\documentclass[twocolumn]{article}
\pdfoutput=1
\usepackage{fullpage}
\usepackage[utf8]{inputenc}
\usepackage{newunicodechar}
\newunicodechar{£}{\sterling}
\newunicodechar{¥}{\yen}
\newunicodechar{¬}{\neg}
\newunicodechar{±}{\pm}
\newunicodechar{·}{\cdotp}
\newunicodechar{×}{\times}
\newunicodechar{ð}{\matheth}
\newunicodechar{÷}{\div}
\newunicodechar{Ƶ}{\Zbar}
\newunicodechar{Α}{\Alpha}
\newunicodechar{Β}{\Beta}
\newunicodechar{Γ}{\Gamma}
\newunicodechar{Δ}{\Delta}
\newunicodechar{Ε}{\Epsilon}
\newunicodechar{Ζ}{\Zeta}
\newunicodechar{Η}{\Eta}
\newunicodechar{Θ}{\Theta}
\newunicodechar{Ι}{\Iota}
\newunicodechar{Κ}{\Kappa}
\newunicodechar{Λ}{\Lambda}
\newunicodechar{Μ}{\Mu}
\newunicodechar{Ν}{\Nu}
\newunicodechar{Ξ}{\Xi}
\newunicodechar{Ο}{\Omicron}
\newunicodechar{Π}{\Pi}
\newunicodechar{Ρ}{\Rho}
\newunicodechar{Σ}{\Sigma}
\newunicodechar{Τ}{\Tau}
\newunicodechar{Υ}{\Upsilon}
\newunicodechar{Φ}{\Phi}
\newunicodechar{Χ}{\Chi}
\newunicodechar{Ψ}{\Psi}
\newunicodechar{Ω}{\Omega}
\newunicodechar{α}{\alpha}
\newunicodechar{β}{\beta}
\newunicodechar{γ}{\gamma}
\newunicodechar{δ}{\delta}
\newunicodechar{ε}{\epsilon}
\newunicodechar{ζ}{\zeta}
\newunicodechar{η}{\eta}
\newunicodechar{θ}{\theta}
\newunicodechar{ι}{\iota}
\newunicodechar{κ}{\kappa}
\newunicodechar{λ}{\lambda}
\newunicodechar{μ}{\mu}
\newunicodechar{ν}{\nu}
\newunicodechar{ξ}{\xi}
\newunicodechar{ο}{\omicron}
\newunicodechar{π}{\pi}
\newunicodechar{ρ}{\rho}
\newunicodechar{ς}{\varsigma}
\newunicodechar{σ}{\sigma}
\newunicodechar{τ}{\tau}
\newunicodechar{υ}{\upsilon}
\newunicodechar{φ}{\varphi}
\newunicodechar{χ}{\chi}
\newunicodechar{ψ}{\psi}
\newunicodechar{ω}{\omega}
\newunicodechar{ϐ}{\varbeta}
\newunicodechar{ϑ}{\vartheta}
\newunicodechar{ϕ}{\phi}
\newunicodechar{ϖ}{\varpi}
\newunicodechar{Ϙ}{\oldKoppa}
\newunicodechar{ϙ}{\oldkoppa}
\newunicodechar{Ϛ}{\Stigma}
\newunicodechar{ϛ}{\stigma}
\newunicodechar{Ϝ}{\Digamma}
\newunicodechar{ϝ}{\digamma}
\newunicodechar{Ϟ}{\Koppa}
\newunicodechar{ϟ}{\koppa}
\newunicodechar{Ϡ}{\Sampi}
\newunicodechar{ϡ}{\sampi}
\newunicodechar{ϰ}{\varkappa}
\newunicodechar{ϱ}{\varrho}
\newunicodechar{ϴ}{\varTheta}
\newunicodechar{ϵ}{\varepsilon}
\newunicodechar{϶}{\backepsilon}
\newunicodechar{―}{\horizbar}
\newunicodechar{‖}{\Vert}
\newunicodechar{‗}{\twolowline}
\newunicodechar{†}{\dagger}
\newunicodechar{‡}{\ddagger}
\newunicodechar{•}{\smblkcircle}
\newunicodechar{‥}{\enleadertwodots}
\newunicodechar{…}{\unicodeellipsis}
\newunicodechar{′}{\prime}
\newunicodechar{″}{\dprime}
\newunicodechar{‴}{\trprime}
\newunicodechar{‵}{\backprime}
\newunicodechar{‶}{\backdprime}
\newunicodechar{‷}{\backtrprime}
\newunicodechar{‸}{\caretinsert}
\newunicodechar{‼}{\Exclam}
\newunicodechar{⁀}{\tieconcat}
\newunicodechar{⁃}{\hyphenbullet}
\newunicodechar{⁄}{\fracslash}
\newunicodechar{⁇}{\Question}
\newunicodechar{⁐}{\closure}
\newunicodechar{⁗}{\qprime}
\newunicodechar{€}{\euro}
\newunicodechar{⃐}{\leftharpoonaccent}
\newunicodechar{⃑}{\rightharpoonaccent}
\newunicodechar{⃒}{\vertoverlay}
\newunicodechar{⃖}{\overleftarrow}
\newunicodechar{⃗}{\vec}
\newunicodechar{⃛}{\dddot}
\newunicodechar{⃜}{\ddddot}
\newunicodechar{⃝}{\enclosecircle}
\newunicodechar{⃞}{\enclosesquare}
\newunicodechar{⃟}{\enclosediamond}
\newunicodechar{⃡}{\overleftrightarrow}
\newunicodechar{⃤}{\enclosetriangle}
\newunicodechar{⃧}{\annuity}
\newunicodechar{ℂ}{\BbbC}
\newunicodechar{ℇ}{\Eulerconst}
\newunicodechar{ℊ}{\mscrg}
\newunicodechar{ℋ}{\mscrH}
\newunicodechar{ℌ}{\mfrakH}
\newunicodechar{ℍ}{\BbbH}
\newunicodechar{ℎ}{\Planckconst}
\newunicodechar{ℏ}{\hslash}
\newunicodechar{ℐ}{\mscrI}
\newunicodechar{ℑ}{\Im}
\newunicodechar{ℒ}{\mscrL}
\newunicodechar{ℓ}{\ell}
\newunicodechar{ℕ}{\BbbN}
\newunicodechar{℘}{\wp}
\newunicodechar{ℙ}{\BbbP}
\newunicodechar{ℚ}{\BbbQ}
\newunicodechar{ℛ}{\mscrR}
\newunicodechar{ℜ}{\Re}
\newunicodechar{ℝ}{\BbbR}
\newunicodechar{ℤ}{\BbbZ}
\newunicodechar{℧}{\mho}
\newunicodechar{ℨ}{\mfrakZ}
\newunicodechar{℩}{\turnediota}
\newunicodechar{Å}{\Angstrom}
\newunicodechar{ℬ}{\mscrB}
\newunicodechar{ℭ}{\mfrakC}
\newunicodechar{ℯ}{\mscre}
\newunicodechar{ℰ}{\mscrE}
\newunicodechar{ℱ}{\mscrF}
\newunicodechar{Ⅎ}{\Finv}
\newunicodechar{ℳ}{\mscrM}
\newunicodechar{ℴ}{\mscro}
\newunicodechar{ℵ}{\aleph}
\newunicodechar{ℶ}{\beth}
\newunicodechar{ℷ}{\gimel}
\newunicodechar{ℸ}{\daleth}
\newunicodechar{ℼ}{\Bbbpi}
\newunicodechar{ℽ}{\Bbbgamma}
\newunicodechar{ℾ}{\BbbGamma}
\newunicodechar{ℿ}{\BbbPi}
\newunicodechar{⅀}{\Bbbsum}
\newunicodechar{⅁}{\Game}
\newunicodechar{⅂}{\sansLturned}
\newunicodechar{⅃}{\sansLmirrored}
\newunicodechar{⅄}{\Yup}
\newunicodechar{ⅅ}{\mitBbbD}
\newunicodechar{ⅆ}{\mitBbbd}
\newunicodechar{ⅇ}{\mitBbbe}
\newunicodechar{ⅈ}{\mitBbbi}
\newunicodechar{ⅉ}{\mitBbbj}
\newunicodechar{⅊}{\PropertyLine}
\newunicodechar{⅋}{\upand} 
\newunicodechar{←}{\leftarrow}
\newunicodechar{↑}{\uparrow}
\newunicodechar{→}{\rightarrow}
\newunicodechar{↓}{\downarrow}
\newunicodechar{↔}{\leftrightarrow}
\newunicodechar{↕}{\updownarrow}
\newunicodechar{↖}{\nwarrow}
\newunicodechar{↗}{\nearrow}
\newunicodechar{↘}{\searrow}
\newunicodechar{↙}{\swarrow}
\newunicodechar{↚}{\nleftarrow}
\newunicodechar{↛}{\nrightarrow}
\newunicodechar{↜}{\leftwavearrow}
\newunicodechar{↝}{\rightwavearrow}
\newunicodechar{↞}{\twoheadleftarrow}
\newunicodechar{↟}{\twoheaduparrow}
\newunicodechar{↠}{\twoheadrightarrow}
\newunicodechar{↡}{\twoheaddownarrow}
\newunicodechar{↢}{\leftarrowtail}
\newunicodechar{↣}{\rightarrowtail}
\newunicodechar{↤}{\mapsfrom}
\newunicodechar{↥}{\mapsup}
\newunicodechar{↦}{\mapsto}
\newunicodechar{↧}{\mapsdown}
\newunicodechar{↨}{\updownarrowbar}
\newunicodechar{↩}{\hookleftarrow}
\newunicodechar{↪}{\hookrightarrow}
\newunicodechar{↫}{\looparrowleft}
\newunicodechar{↬}{\looparrowright}
\newunicodechar{↭}{\leftrightsquigarrow}
\newunicodechar{↮}{\nleftrightarrow}
\newunicodechar{↯}{\downzigzagarrow}
\newunicodechar{↰}{\Lsh}
\newunicodechar{↱}{\Rsh}
\newunicodechar{↲}{\Ldsh}
\newunicodechar{↳}{\Rdsh}
\newunicodechar{↴}{\linefeed}
\newunicodechar{↵}{\carriagereturn}
\newunicodechar{↶}{\curvearrowleft}
\newunicodechar{↷}{\curvearrowright}
\newunicodechar{↸}{\barovernorthwestarrow}
\newunicodechar{↹}{\barleftarrowrightarrowbar}
\newunicodechar{↺}{\acwopencirclearrow}
\newunicodechar{↻}{\cwopencirclearrow}
\newunicodechar{↼}{\leftharpoonup}
\newunicodechar{↽}{\leftharpoondown}
\newunicodechar{↾}{\upharpoonright}
\newunicodechar{↿}{\upharpoonleft}
\newunicodechar{⇀}{\rightharpoonup}
\newunicodechar{⇁}{\rightharpoondown}
\newunicodechar{⇂}{\downharpoonright}
\newunicodechar{⇃}{\downharpoonleft}
\newunicodechar{⇄}{\rightleftarrows}
\newunicodechar{⇅}{\updownarrows}
\newunicodechar{⇆}{\leftrightarrows}
\newunicodechar{⇇}{\leftleftarrows}
\newunicodechar{⇈}{\upuparrows}
\newunicodechar{⇉}{\rightrightarrows}
\newunicodechar{⇊}{\downdownarrows}
\newunicodechar{⇋}{\leftrightharpoons}
\newunicodechar{⇌}{\rightleftharpoons}
\newunicodechar{⇍}{\nLeftarrow}
\newunicodechar{⇎}{\nLeftrightarrow}
\newunicodechar{⇏}{\nRightarrow}
\newunicodechar{⇐}{\Leftarrow}
\newunicodechar{⇑}{\Uparrow}
\newunicodechar{⇒}{\Rightarrow}
\newunicodechar{⇓}{\Downarrow}
\newunicodechar{⇔}{\Leftrightarrow}
\newunicodechar{⇕}{\Updownarrow}
\newunicodechar{⇖}{\Nwarrow}
\newunicodechar{⇗}{\Nearrow}
\newunicodechar{⇘}{\Searrow}
\newunicodechar{⇙}{\Swarrow}
\newunicodechar{⇚}{\Lleftarrow}
\newunicodechar{⇛}{\Rrightarrow}
\newunicodechar{⇜}{\leftsquigarrow}
\newunicodechar{⇝}{\rightsquigarrow}
\newunicodechar{⇞}{\nHuparrow}
\newunicodechar{⇟}{\nHdownarrow}
\newunicodechar{⇠}{\leftdasharrow}
\newunicodechar{⇡}{\updasharrow}
\newunicodechar{⇢}{\rightdasharrow}
\newunicodechar{⇣}{\downdasharrow}
\newunicodechar{⇤}{\barleftarrow}
\newunicodechar{⇥}{\rightarrowbar}
\newunicodechar{⇦}{\leftwhitearrow}
\newunicodechar{⇧}{\upwhitearrow}
\newunicodechar{⇨}{\rightwhitearrow}
\newunicodechar{⇩}{\downwhitearrow}
\newunicodechar{⇪}{\whitearrowupfrombar}
\newunicodechar{⇴}{\circleonrightarrow}
\newunicodechar{⇵}{\downuparrows}
\newunicodechar{⇶}{\rightthreearrows}
\newunicodechar{⇷}{\nvleftarrow}
\newunicodechar{⇸}{\nvrightarrow}
\newunicodechar{⇹}{\nvleftrightarrow}
\newunicodechar{⇺}{\nVleftarrow}
\newunicodechar{⇻}{\nVrightarrow}
\newunicodechar{⇼}{\nVleftrightarrow}
\newunicodechar{⇽}{\leftarrowtriangle}
\newunicodechar{⇾}{\rightarrowtriangle}
\newunicodechar{⇿}{\leftrightarrowtriangle}
\newunicodechar{∀}{\forall}
\newunicodechar{∁}{\complement}
\newunicodechar{∂}{\partial}
\newunicodechar{∃}{\exists}
\newunicodechar{∄}{\nexists}
\newunicodechar{∅}{\varnothing}
\newunicodechar{∆}{\increment}
\newunicodechar{∇}{\nabla}
\newunicodechar{∈}{\in}
\newunicodechar{∉}{\notin}
\newunicodechar{∊}{\smallin}
\newunicodechar{∋}{\ni}
\newunicodechar{∌}{\nni}
\newunicodechar{∍}{\smallni}
\newunicodechar{∎}{\QED}
\newunicodechar{∏}{\prod}
\newunicodechar{∐}{\coprod}
\newunicodechar{∑}{\sum}
\newunicodechar{−}{\minus}
\newunicodechar{∓}{\mp}
\newunicodechar{∔}{\dotplus}
\newunicodechar{∕}{\divslash}
\newunicodechar{∖}{\smallsetminus}
\newunicodechar{∗}{\ast}
\newunicodechar{∘}{\vysmwhtcircle}
\newunicodechar{∙}{\vysmblkcircle}
\newunicodechar{√}{\sqrt}
\newunicodechar{∛}{\cuberoot}
\newunicodechar{∜}{\fourthroot}
\newunicodechar{∝}{\propto}
\newunicodechar{∞}{\infty}
\newunicodechar{∟}{\rightangle}
\newunicodechar{∠}{\angle}
\newunicodechar{∡}{\measuredangle}
\newunicodechar{∢}{\sphericalangle}
\newunicodechar{∣}{\mid}
\newunicodechar{∤}{\nmid}
\newunicodechar{∥}{\parallel}
\newunicodechar{∦}{\nparallel}
\newunicodechar{∧}{\wedge}
\newunicodechar{∨}{\vee}
\newunicodechar{∩}{\cap}
\newunicodechar{∪}{\cup}
\newunicodechar{∫}{\int}
\newunicodechar{∬}{\iint}
\newunicodechar{∭}{\iiint}
\newunicodechar{∮}{\oint}
\newunicodechar{∯}{\oiint}
\newunicodechar{∰}{\oiiint}
\newunicodechar{∱}{\intclockwise}
\newunicodechar{∲}{\varointclockwise}
\newunicodechar{∳}{\ointctrclockwise}
\newunicodechar{∴}{\therefore}
\newunicodechar{∵}{\because}
\newunicodechar{∶}{\mathratio}
\newunicodechar{∷}{\Colon}
\newunicodechar{∸}{\dotminus}
\newunicodechar{∹}{\dashcolon}
\newunicodechar{∺}{\dotsminusdots}
\newunicodechar{∻}{\kernelcontraction}
\newunicodechar{∼}{\sim}
\newunicodechar{∽}{\backsim}
\newunicodechar{∾}{\invlazys}
\newunicodechar{∿}{\sinewave}
\newunicodechar{≀}{\wr}
\newunicodechar{≁}{\nsim}
\newunicodechar{≂}{\eqsim}
\newunicodechar{≃}{\simeq}
\newunicodechar{≄}{\nsime}
\newunicodechar{≅}{\cong}
\newunicodechar{≆}{\simneqq}
\newunicodechar{≇}{\ncong}
\newunicodechar{≈}{\approx}
\newunicodechar{≉}{\napprox}
\newunicodechar{≊}{\approxeq}
\newunicodechar{≋}{\approxident}
\newunicodechar{≌}{\backcong}
\newunicodechar{≍}{\asymp}
\newunicodechar{≎}{\Bumpeq}
\newunicodechar{≏}{\bumpeq}
\newunicodechar{≐}{\doteq}
\newunicodechar{≑}{\Doteq}
\newunicodechar{≒}{\fallingdotseq}
\newunicodechar{≓}{\risingdotseq}
\newunicodechar{≔}{\coloneq}
\newunicodechar{≕}{\eqcolon}
\newunicodechar{≖}{\eqcirc}
\newunicodechar{≗}{\circeq}
\newunicodechar{≘}{\arceq}
\newunicodechar{≙}{\wedgeq}
\newunicodechar{≚}{\veeeq}
\newunicodechar{≛}{\stareq}
\newunicodechar{≜}{\triangleq}
\newunicodechar{≝}{\eqdef}
\newunicodechar{≞}{\measeq}
\newunicodechar{≟}{\questeq}
\newunicodechar{≠}{\ne}
\newunicodechar{≡}{\equiv}
\newunicodechar{≢}{\nequiv}
\newunicodechar{≣}{\Equiv}
\newunicodechar{≤}{\leq}
\newunicodechar{≥}{\geq}
\newunicodechar{≦}{\leqq}
\newunicodechar{≧}{\geqq}
\newunicodechar{≨}{\lneqq}
\newunicodechar{≩}{\gneqq}
\newunicodechar{≪}{\ll}
\newunicodechar{≫}{\gg}
\newunicodechar{≬}{\between}
\newunicodechar{≭}{\nasymp}
\newunicodechar{≮}{\nless}
\newunicodechar{≯}{\ngtr}
\newunicodechar{≰}{\nleq}
\newunicodechar{≱}{\ngeq}
\newunicodechar{≲}{\lesssim}
\newunicodechar{≳}{\gtrsim}
\newunicodechar{≴}{\nlesssim}
\newunicodechar{≵}{\ngtrsim}
\newunicodechar{≶}{\lessgtr}
\newunicodechar{≷}{\gtrless}
\newunicodechar{≸}{\nlessgtr}
\newunicodechar{≹}{\ngtrless}
\newunicodechar{≺}{\prec}
\newunicodechar{≻}{\succ}
\newunicodechar{≼}{\preccurlyeq}
\newunicodechar{≽}{\succcurlyeq}
\newunicodechar{≾}{\precsim}
\newunicodechar{≿}{\succsim}
\newunicodechar{⊀}{\nprec}
\newunicodechar{⊁}{\nsucc}
\newunicodechar{⊂}{\subset}
\newunicodechar{⊃}{\supset}
\newunicodechar{⊄}{\nsubset}
\newunicodechar{⊅}{\nsupset}
\newunicodechar{⊆}{\subseteq}
\newunicodechar{⊇}{\supseteq}
\newunicodechar{⊈}{\nsubseteq}
\newunicodechar{⊉}{\nsupseteq}
\newunicodechar{⊊}{\subsetneq}
\newunicodechar{⊋}{\supsetneq}
\newunicodechar{⊌}{\cupleftarrow}
\newunicodechar{⊍}{\cupdot}
\newunicodechar{⊎}{\uplus}
\newunicodechar{⊏}{\sqsubset}
\newunicodechar{⊐}{\sqsupset}
\newunicodechar{⊑}{\sqsubseteq}
\newunicodechar{⊒}{\sqsupseteq}
\newunicodechar{⊓}{\sqcap}
\newunicodechar{⊔}{\sqcup}
\newunicodechar{⊕}{\oplus}
\newunicodechar{⊖}{\ominus}
\newunicodechar{⊗}{\otimes}
\newunicodechar{⊘}{\oslash}
\newunicodechar{⊙}{\odot}
\newunicodechar{⊚}{\circledcirc}
\newunicodechar{⊛}{\circledast}
\newunicodechar{⊜}{\circledequal}
\newunicodechar{⊝}{\circleddash}
\newunicodechar{⊞}{\boxplus}
\newunicodechar{⊟}{\boxminus}
\newunicodechar{⊠}{\boxtimes}
\newunicodechar{⊡}{\boxdot}
\newunicodechar{⊢}{\vdash}
\newunicodechar{⊣}{\dashv}
\newunicodechar{⊤}{\top}
\newunicodechar{⊥}{\bot}
\newunicodechar{⊦}{\assert}
\newunicodechar{⊧}{\models}
\newunicodechar{⊨}{\vDash}
\newunicodechar{⊩}{\Vdash}
\newunicodechar{⊪}{\Vvdash}
\newunicodechar{⊫}{\VDash}
\newunicodechar{⊬}{\nvdash}
\newunicodechar{⊭}{\nvDash}
\newunicodechar{⊮}{\nVdash}
\newunicodechar{⊯}{\nVDash}
\newunicodechar{⊰}{\prurel}
\newunicodechar{⊱}{\scurel}
\newunicodechar{⊲}{\vartriangleleft}
\newunicodechar{⊳}{\vartriangleright}
\newunicodechar{⊴}{\trianglelefteq}
\newunicodechar{⊵}{\trianglerighteq}
\newunicodechar{⊶}{\origof}
\newunicodechar{⊷}{\imageof}
\newunicodechar{⊸}{\multimap}
\newunicodechar{⊹}{\hermitmatrix}
\newunicodechar{⊺}{\intercal}
\newunicodechar{⊻}{\veebar}
\newunicodechar{⊼}{\barwedge}
\newunicodechar{⊽}{\barvee}
\newunicodechar{⊾}{\measuredrightangle}
\newunicodechar{⊿}{\varlrtriangle}
\newunicodechar{⋀}{\bigwedge}
\newunicodechar{⋁}{\bigvee}
\newunicodechar{⋂}{\bigcap}
\newunicodechar{⋃}{\bigcup}
\newunicodechar{⋄}{\smwhtdiamond}
\newunicodechar{⋅}{\cdot}
\newunicodechar{⋆}{\star}
\newunicodechar{⋇}{\divideontimes}
\newunicodechar{⋈}{\bowtie}
\newunicodechar{⋉}{\ltimes}
\newunicodechar{⋊}{\rtimes}
\newunicodechar{⋋}{\leftthreetimes}
\newunicodechar{⋌}{\rightthreetimes}
\newunicodechar{⋍}{\backsimeq}
\newunicodechar{⋎}{\curlyvee}
\newunicodechar{⋏}{\curlywedge}
\newunicodechar{⋐}{\Subset}
\newunicodechar{⋑}{\Supset}
\newunicodechar{⋒}{\Cap}
\newunicodechar{⋓}{\Cup}
\newunicodechar{⋔}{\pitchfork}
\newunicodechar{⋕}{\equalparallel}
\newunicodechar{⋖}{\lessdot}
\newunicodechar{⋗}{\gtrdot}
\newunicodechar{⋘}{\lll}
\newunicodechar{⋙}{\ggg}
\newunicodechar{⋚}{\lesseqgtr}
\newunicodechar{⋛}{\gtreqless}
\newunicodechar{⋜}{\eqless}
\newunicodechar{⋝}{\eqgtr}
\newunicodechar{⋞}{\curlyeqprec}
\newunicodechar{⋟}{\curlyeqsucc}
\newunicodechar{⋠}{\npreccurlyeq}
\newunicodechar{⋡}{\nsucccurlyeq}
\newunicodechar{⋢}{\nsqsubseteq}
\newunicodechar{⋣}{\nsqsupseteq}
\newunicodechar{⋤}{\sqsubsetneq}
\newunicodechar{⋥}{\sqsupsetneq}
\newunicodechar{⋦}{\lnsim}
\newunicodechar{⋧}{\gnsim}
\newunicodechar{⋨}{\precnsim}
\newunicodechar{⋩}{\succnsim}
\newunicodechar{⋪}{\nvartriangleleft}
\newunicodechar{⋫}{\nvartriangleright}
\newunicodechar{⋬}{\ntrianglelefteq}
\newunicodechar{⋭}{\ntrianglerighteq}
\newunicodechar{⋮}{\vdots}
\newunicodechar{⋯}{\unicodecdots}
\newunicodechar{⋰}{\adots}
\newunicodechar{⋱}{\ddots}
\newunicodechar{⋲}{\disin}
\newunicodechar{⋳}{\varisins}
\newunicodechar{⋴}{\isins}
\newunicodechar{⋵}{\isindot}
\newunicodechar{⋶}{\varisinobar}
\newunicodechar{⋷}{\isinobar}
\newunicodechar{⋸}{\isinvb}
\newunicodechar{⋹}{\isinE}
\newunicodechar{⋺}{\nisd}
\newunicodechar{⋻}{\varnis}
\newunicodechar{⋼}{\nis}
\newunicodechar{⋽}{\varniobar}
\newunicodechar{⋾}{\niobar}
\newunicodechar{⋿}{\bagmember}
\newunicodechar{⌀}{\diameter}
\newunicodechar{⌂}{\house}
\newunicodechar{⌅}{\varbarwedge}
\newunicodechar{⌆}{\vardoublebarwedge}
\newunicodechar{⌈}{\lceil}
\newunicodechar{⌉}{\rceil}
\newunicodechar{⌊}{\lfloor}
\newunicodechar{⌋}{\rfloor}
\newunicodechar{⌐}{\invnot}
\newunicodechar{⌑}{\sqlozenge}
\newunicodechar{⌒}{\profline}
\newunicodechar{⌓}{\profsurf}
\newunicodechar{⌗}{\viewdata}
\newunicodechar{⌙}{\turnednot}
\newunicodechar{⌜}{\ulcorner}
\newunicodechar{⌝}{\urcorner}
\newunicodechar{⌞}{\llcorner}
\newunicodechar{⌟}{\lrcorner}
\newunicodechar{⌠}{\inttop}
\newunicodechar{⌡}{\intbottom}
\newunicodechar{⌢}{\frown}
\newunicodechar{⌣}{\smile}
\newunicodechar{⌬}{\varhexagonlrbonds}
\newunicodechar{⌲}{\conictaper}
\newunicodechar{⌶}{\topbot}
\newunicodechar{⌽}{\obar}
\newunicodechar{⌿}{\APLnotslash}
\newunicodechar{⍀}{\APLnotbackslash}
\newunicodechar{⍓}{\APLboxupcaret}
\newunicodechar{⍰}{\APLboxquestion}
\newunicodechar{⍼}{\rangledownzigzagarrow}
\newunicodechar{⎔}{\hexagon}
\newunicodechar{⎛}{\lparenuend}
\newunicodechar{⎜}{\lparenextender}
\newunicodechar{⎝}{\lparenlend}
\newunicodechar{⎞}{\rparenuend}
\newunicodechar{⎟}{\rparenextender}
\newunicodechar{⎠}{\rparenlend}
\newunicodechar{⎡}{\lbrackuend}
\newunicodechar{⎢}{\lbrackextender}
\newunicodechar{⎣}{\lbracklend}
\newunicodechar{⎤}{\rbrackuend}
\newunicodechar{⎥}{\rbrackextender}
\newunicodechar{⎦}{\rbracklend}
\newunicodechar{⎧}{\lbraceuend}
\newunicodechar{⎨}{\lbracemid}
\newunicodechar{⎩}{\lbracelend}
\newunicodechar{⎪}{\vbraceextender}
\newunicodechar{⎫}{\rbraceuend}
\newunicodechar{⎬}{\rbracemid}
\newunicodechar{⎭}{\rbracelend}
\newunicodechar{⎮}{\intextender}
\newunicodechar{⎯}{\harrowextender}
\newunicodechar{⎰}{\lmoustache}
\newunicodechar{⎱}{\rmoustache}
\newunicodechar{⎲}{\sumtop}
\newunicodechar{⎳}{\sumbottom}
\newunicodechar{⎴}{\overbracket}
\newunicodechar{⎵}{\underbracket}
\newunicodechar{⎶}{\bbrktbrk}
\newunicodechar{⎷}{\sqrtbottom}
\newunicodechar{⎸}{\lvboxline}
\newunicodechar{⎹}{\rvboxline}
\newunicodechar{⏎}{\varcarriagereturn}
\newunicodechar{⏜}{\overparen}
\newunicodechar{⏝}{\underparen}
\newunicodechar{⏞}{\overbrace}
\newunicodechar{⏟}{\underbrace}
\newunicodechar{⏠}{\obrbrak}
\newunicodechar{⏡}{\ubrbrak}
\newunicodechar{⏢}{\trapezium}
\newunicodechar{⏣}{\benzenr}
\newunicodechar{⏤}{\strns}
\newunicodechar{⏥}{\fltns}
\newunicodechar{⏦}{\accurrent}
\newunicodechar{⏧}{\elinters}
\newunicodechar{␢}{\blanksymbol}
\newunicodechar{␣}{\mathvisiblespace}
\newunicodechar{┆}{\bdtriplevdash}
\newunicodechar{▀}{\blockuphalf}
\newunicodechar{▄}{\blocklowhalf}
\newunicodechar{█}{\blockfull}
\newunicodechar{▌}{\blocklefthalf}
\newunicodechar{▐}{\blockrighthalf}
\newunicodechar{░}{\blockqtrshaded}
\newunicodechar{▒}{\blockhalfshaded}
\newunicodechar{▓}{\blockthreeqtrshaded}
\newunicodechar{■}{\mdlgblksquare}
\newunicodechar{□}{\mdlgwhtsquare}
\newunicodechar{▢}{\squoval}
\newunicodechar{▣}{\blackinwhitesquare}
\newunicodechar{▤}{\squarehfill}
\newunicodechar{▥}{\squarevfill}
\newunicodechar{▦}{\squarehvfill}
\newunicodechar{▧}{\squarenwsefill}
\newunicodechar{▨}{\squareneswfill}
\newunicodechar{▩}{\squarecrossfill}
\newunicodechar{▪}{\smblksquare}
\newunicodechar{▫}{\smwhtsquare}
\newunicodechar{▬}{\hrectangleblack}
\newunicodechar{▭}{\hrectangle}
\newunicodechar{▮}{\vrectangleblack}
\newunicodechar{▯}{\vrectangle}
\newunicodechar{▰}{\parallelogramblack}
\newunicodechar{▱}{\parallelogram}
\newunicodechar{▲}{\bigblacktriangleup}
\newunicodechar{△}{\bigtriangleup}
\newunicodechar{▴}{\blacktriangle}
\newunicodechar{▵}{\vartriangle}
\newunicodechar{▶}{\blacktriangleright}
\newunicodechar{▷}{\triangleright}
\newunicodechar{▸}{\smallblacktriangleright}
\newunicodechar{▹}{\smalltriangleright}
\newunicodechar{►}{\blackpointerright}
\newunicodechar{▻}{\whitepointerright}
\newunicodechar{▼}{\bigblacktriangledown}
\newunicodechar{▽}{\bigtriangledown}
\newunicodechar{▾}{\blacktriangledown}
\newunicodechar{▿}{\triangledown}
\newunicodechar{◀}{\blacktriangleleft}
\newunicodechar{◁}{\triangleleft}
\newunicodechar{◂}{\smallblacktriangleleft}
\newunicodechar{◃}{\smalltriangleleft}
\newunicodechar{◄}{\blackpointerleft}
\newunicodechar{◅}{\whitepointerleft}
\newunicodechar{◆}{\mdlgblkdiamond}
\newunicodechar{◇}{\mdlgwhtdiamond}
\newunicodechar{◈}{\blackinwhitediamond}
\newunicodechar{◉}{\fisheye}
\newunicodechar{◊}{\mdlgwhtlozenge}
\newunicodechar{○}{\mdlgwhtcircle}
\newunicodechar{◌}{\dottedcircle}
\newunicodechar{◍}{\circlevertfill}
\newunicodechar{◎}{\bullseye}
\newunicodechar{●}{\mdlgblkcircle}
\newunicodechar{◐}{\circlelefthalfblack}
\newunicodechar{◑}{\circlerighthalfblack}
\newunicodechar{◒}{\circlebottomhalfblack}
\newunicodechar{◓}{\circletophalfblack}
\newunicodechar{◔}{\circleurquadblack}
\newunicodechar{◕}{\blackcircleulquadwhite}
\newunicodechar{◖}{\blacklefthalfcircle}
\newunicodechar{◗}{\blackrighthalfcircle}
\newunicodechar{◘}{\inversebullet}
\newunicodechar{◙}{\inversewhitecircle}
\newunicodechar{◚}{\invwhiteupperhalfcircle}
\newunicodechar{◛}{\invwhitelowerhalfcircle}
\newunicodechar{◜}{\ularc}
\newunicodechar{◝}{\urarc}
\newunicodechar{◞}{\lrarc}
\newunicodechar{◟}{\llarc}
\newunicodechar{◠}{\topsemicircle}
\newunicodechar{◡}{\botsemicircle}
\newunicodechar{◢}{\lrblacktriangle}
\newunicodechar{◣}{\llblacktriangle}
\newunicodechar{◤}{\ulblacktriangle}
\newunicodechar{◥}{\urblacktriangle}
\newunicodechar{◦}{\smwhtcircle}
\newunicodechar{◧}{\squareleftblack}
\newunicodechar{◨}{\squarerightblack}
\newunicodechar{◩}{\squareulblack}
\newunicodechar{◪}{\squarelrblack}
\newunicodechar{◫}{\boxbar}
\newunicodechar{◬}{\trianglecdot}
\newunicodechar{◭}{\triangleleftblack}
\newunicodechar{◮}{\trianglerightblack}
\newunicodechar{◯}{\lgwhtcircle}
\newunicodechar{◰}{\squareulquad}
\newunicodechar{◱}{\squarellquad}
\newunicodechar{◲}{\squarelrquad}
\newunicodechar{◳}{\squareurquad}
\newunicodechar{◴}{\circleulquad}
\newunicodechar{◵}{\circlellquad}
\newunicodechar{◶}{\circlelrquad}
\newunicodechar{◷}{\circleurquad}
\newunicodechar{◸}{\ultriangle}
\newunicodechar{◹}{\urtriangle}
\newunicodechar{◺}{\lltriangle}
\newunicodechar{◻}{\mdwhtsquare}
\newunicodechar{◼}{\mdblksquare}
\newunicodechar{◽}{\mdsmwhtsquare}
\newunicodechar{◾}{\mdsmblksquare}
\newunicodechar{◿}{\lrtriangle}
\newunicodechar{★}{\bigstar}
\newunicodechar{☆}{\bigwhitestar}
\newunicodechar{☉}{\astrosun}
\newunicodechar{☡}{\danger}
\newunicodechar{☻}{\blacksmiley}
\newunicodechar{☼}{\sun}
\newunicodechar{☽}{\rightmoon}
\newunicodechar{☾}{\leftmoon}
\newunicodechar{♀}{\female}
\newunicodechar{♂}{\male}
\newunicodechar{♠}{\spadesuit}
\newunicodechar{♡}{\heartsuit}
\newunicodechar{♢}{\diamondsuit}
\newunicodechar{♣}{\clubsuit}
\newunicodechar{♤}{\varspadesuit}
\newunicodechar{♥}{\varheartsuit}
\newunicodechar{♦}{\vardiamondsuit}
\newunicodechar{♧}{\varclubsuit}
\newunicodechar{♩}{\quarternote}
\newunicodechar{♪}{\eighthnote}
\newunicodechar{♫}{\twonotes}
\newunicodechar{♭}{\flat}
\newunicodechar{♮}{\natural}
\newunicodechar{♯}{\sharp}
\newunicodechar{♾}{\acidfree}
\newunicodechar{⚀}{\dicei}
\newunicodechar{⚁}{\diceii}
\newunicodechar{⚂}{\diceiii}
\newunicodechar{⚃}{\diceiv}
\newunicodechar{⚄}{\dicev}
\newunicodechar{⚅}{\dicevi}
\newunicodechar{⚆}{\circledrightdot}
\newunicodechar{⚇}{\circledtwodots}
\newunicodechar{⚈}{\blackcircledrightdot}
\newunicodechar{⚉}{\blackcircledtwodots}
\newunicodechar{⚥}{\Hermaphrodite}
\newunicodechar{⚪}{\mdwhtcircle}
\newunicodechar{⚫}{\mdblkcircle}
\newunicodechar{⚬}{\mdsmwhtcircle}
\newunicodechar{⚲}{\neuter}
\newunicodechar{✓}{\checkmark}
\newunicodechar{✠}{\maltese}
\newunicodechar{✪}{\circledstar}
\newunicodechar{✶}{\varstar}
\newunicodechar{✽}{\dingasterisk}
\newunicodechar{❲}{\lbrbrak}
\newunicodechar{❳}{\rbrbrak}
\newunicodechar{➛}{\draftingarrow}
\newunicodechar{⟀}{\threedangle}
\newunicodechar{⟁}{\whiteinwhitetriangle}
\newunicodechar{⟂}{\perp}
\newunicodechar{⟃}{\subsetcirc}
\newunicodechar{⟄}{\supsetcirc}
\newunicodechar{⟅}{\lbag}
\newunicodechar{⟆}{\rbag}
\newunicodechar{⟇}{\veedot}
\newunicodechar{⟈}{\bsolhsub}
\newunicodechar{⟉}{\suphsol}
\newunicodechar{⟌}{\longdivision}
\newunicodechar{⟐}{\diamondcdot}
\newunicodechar{⟑}{\wedgedot}
\newunicodechar{⟒}{\upin}
\newunicodechar{⟓}{\pullback}
\newunicodechar{⟔}{\pushout}
\newunicodechar{⟕}{\leftouterjoin}
\newunicodechar{⟖}{\rightouterjoin}
\newunicodechar{⟗}{\fullouterjoin}
\newunicodechar{⟘}{\bigbot}
\newunicodechar{⟙}{\bigtop}
\newunicodechar{⟚}{\DashVDash}
\newunicodechar{⟛}{\dashVdash}
\newunicodechar{⟜}{\multimapinv}
\newunicodechar{⟝}{\vlongdash}
\newunicodechar{⟞}{\longdashv}
\newunicodechar{⟟}{\cirbot}
\newunicodechar{⟠}{\lozengeminus}
\newunicodechar{⟡}{\concavediamond}
\newunicodechar{⟢}{\concavediamondtickleft}
\newunicodechar{⟣}{\concavediamondtickright}
\newunicodechar{⟤}{\whitesquaretickleft}
\newunicodechar{⟥}{\whitesquaretickright}
\newunicodechar{⟦}{\lBrack}
\newunicodechar{⟧}{\rBrack}
\newunicodechar{⟨}{\langle}
\newunicodechar{⟩}{\rangle}
\newunicodechar{⟪}{\lAngle}
\newunicodechar{⟫}{\rAngle}
\newunicodechar{⟬}{\Lbrbrak}
\newunicodechar{⟭}{\Rbrbrak}
\newunicodechar{⟮}{\lgroup}
\newunicodechar{⟯}{\rgroup}
\newunicodechar{⟰}{\UUparrow}
\newunicodechar{⟱}{\DDownarrow}
\newunicodechar{⟲}{\acwgapcirclearrow}
\newunicodechar{⟳}{\cwgapcirclearrow}
\newunicodechar{⟴}{\rightarrowonoplus}
\newunicodechar{⟵}{\longleftarrow}
\newunicodechar{⟶}{\longrightarrow}
\newunicodechar{⟷}{\longleftrightarrow}
\newunicodechar{⟸}{\Longleftarrow}
\newunicodechar{⟹}{\Longrightarrow}
\newunicodechar{⟺}{\Longleftrightarrow}
\newunicodechar{⟻}{\longmapsfrom}
\newunicodechar{⟼}{\longmapsto}
\newunicodechar{⟽}{\Longmapsfrom}
\newunicodechar{⟾}{\Longmapsto}
\newunicodechar{⟿}{\longrightsquigarrow}
\newunicodechar{⤀}{\nvtwoheadrightarrow}
\newunicodechar{⤁}{\nVtwoheadrightarrow}
\newunicodechar{⤂}{\nvLeftarrow}
\newunicodechar{⤃}{\nvRightarrow}
\newunicodechar{⤄}{\nvLeftrightarrow}
\newunicodechar{⤅}{\twoheadmapsto}
\newunicodechar{⤆}{\Mapsfrom}
\newunicodechar{⤇}{\Mapsto}
\newunicodechar{⤈}{\downarrowbarred}
\newunicodechar{⤉}{\uparrowbarred}
\newunicodechar{⤊}{\Uuparrow}
\newunicodechar{⤋}{\Ddownarrow}
\newunicodechar{⤌}{\leftbkarrow}
\newunicodechar{⤍}{\rightbkarrow}
\newunicodechar{⤎}{\leftdbkarrow}
\newunicodechar{⤏}{\dbkarow}
\newunicodechar{⤐}{\drbkarow}
\newunicodechar{⤑}{\rightdotarrow}
\newunicodechar{⤒}{\baruparrow}
\newunicodechar{⤓}{\downarrowbar}
\newunicodechar{⤔}{\nvrightarrowtail}
\newunicodechar{⤕}{\nVrightarrowtail}
\newunicodechar{⤖}{\twoheadrightarrowtail}
\newunicodechar{⤗}{\nvtwoheadrightarrowtail}
\newunicodechar{⤘}{\nVtwoheadrightarrowtail}
\newunicodechar{⤙}{\lefttail}
\newunicodechar{⤚}{\righttail}
\newunicodechar{⤛}{\leftdbltail}
\newunicodechar{⤜}{\rightdbltail}
\newunicodechar{⤝}{\diamondleftarrow}
\newunicodechar{⤞}{\rightarrowdiamond}
\newunicodechar{⤟}{\diamondleftarrowbar}
\newunicodechar{⤠}{\barrightarrowdiamond}
\newunicodechar{⤡}{\nwsearrow}
\newunicodechar{⤢}{\neswarrow}
\newunicodechar{⤣}{\hknwarrow}
\newunicodechar{⤤}{\hknearrow}
\newunicodechar{⤥}{\hksearow}
\newunicodechar{⤦}{\hkswarow}
\newunicodechar{⤧}{\tona}
\newunicodechar{⤨}{\toea}
\newunicodechar{⤩}{\tosa}
\newunicodechar{⤪}{\towa}
\newunicodechar{⤫}{\rdiagovfdiag}
\newunicodechar{⤬}{\fdiagovrdiag}
\newunicodechar{⤭}{\seovnearrow}
\newunicodechar{⤮}{\neovsearrow}
\newunicodechar{⤯}{\fdiagovnearrow}
\newunicodechar{⤰}{\rdiagovsearrow}
\newunicodechar{⤱}{\neovnwarrow}
\newunicodechar{⤲}{\nwovnearrow}
\newunicodechar{⤳}{\rightcurvedarrow}
\newunicodechar{⤴}{\uprightcurvearrow}
\newunicodechar{⤵}{\downrightcurvedarrow}
\newunicodechar{⤶}{\leftdowncurvedarrow}
\newunicodechar{⤷}{\rightdowncurvedarrow}
\newunicodechar{⤸}{\cwrightarcarrow}
\newunicodechar{⤹}{\acwleftarcarrow}
\newunicodechar{⤺}{\acwoverarcarrow}
\newunicodechar{⤻}{\acwunderarcarrow}
\newunicodechar{⤼}{\curvearrowrightminus}
\newunicodechar{⤽}{\curvearrowleftplus}
\newunicodechar{⤾}{\cwundercurvearrow}
\newunicodechar{⤿}{\ccwundercurvearrow}
\newunicodechar{⥀}{\acwcirclearrow}
\newunicodechar{⥁}{\cwcirclearrow}
\newunicodechar{⥂}{\rightarrowshortleftarrow}
\newunicodechar{⥃}{\leftarrowshortrightarrow}
\newunicodechar{⥄}{\shortrightarrowleftarrow}
\newunicodechar{⥅}{\rightarrowplus}
\newunicodechar{⥆}{\leftarrowplus}
\newunicodechar{⥇}{\rightarrowx}
\newunicodechar{⥈}{\leftrightarrowcircle}
\newunicodechar{⥉}{\twoheaduparrowcircle}
\newunicodechar{⥊}{\leftrightharpoonupdown}
\newunicodechar{⥋}{\leftrightharpoondownup}
\newunicodechar{⥌}{\updownharpoonrightleft}
\newunicodechar{⥍}{\updownharpoonleftright}
\newunicodechar{⥎}{\leftrightharpoonupup}
\newunicodechar{⥏}{\updownharpoonrightright}
\newunicodechar{⥐}{\leftrightharpoondowndown}
\newunicodechar{⥑}{\updownharpoonleftleft}
\newunicodechar{⥒}{\barleftharpoonup}
\newunicodechar{⥓}{\rightharpoonupbar}
\newunicodechar{⥔}{\barupharpoonright}
\newunicodechar{⥕}{\downharpoonrightbar}
\newunicodechar{⥖}{\barleftharpoondown}
\newunicodechar{⥗}{\rightharpoondownbar}
\newunicodechar{⥘}{\barupharpoonleft}
\newunicodechar{⥙}{\downharpoonleftbar}
\newunicodechar{⥚}{\leftharpoonupbar}
\newunicodechar{⥛}{\barrightharpoonup}
\newunicodechar{⥜}{\upharpoonrightbar}
\newunicodechar{⥝}{\bardownharpoonright}
\newunicodechar{⥞}{\leftharpoondownbar}
\newunicodechar{⥟}{\barrightharpoondown}
\newunicodechar{⥠}{\upharpoonleftbar}
\newunicodechar{⥡}{\bardownharpoonleft}
\newunicodechar{⥢}{\leftharpoonsupdown}
\newunicodechar{⥣}{\upharpoonsleftright}
\newunicodechar{⥤}{\rightharpoonsupdown}
\newunicodechar{⥥}{\downharpoonsleftright}
\newunicodechar{⥦}{\leftrightharpoonsup}
\newunicodechar{⥧}{\leftrightharpoonsdown}
\newunicodechar{⥨}{\rightleftharpoonsup}
\newunicodechar{⥩}{\rightleftharpoonsdown}
\newunicodechar{⥪}{\leftharpoonupdash}
\newunicodechar{⥫}{\dashleftharpoondown}
\newunicodechar{⥬}{\rightharpoonupdash}
\newunicodechar{⥭}{\dashrightharpoondown}
\newunicodechar{⥮}{\updownharpoonsleftright}
\newunicodechar{⥯}{\downupharpoonsleftright}
\newunicodechar{⥰}{\rightimply}
\newunicodechar{⥱}{\equalrightarrow}
\newunicodechar{⥲}{\similarrightarrow}
\newunicodechar{⥳}{\leftarrowsimilar}
\newunicodechar{⥴}{\rightarrowsimilar}
\newunicodechar{⥵}{\rightarrowapprox}
\newunicodechar{⥶}{\ltlarr}
\newunicodechar{⥷}{\leftarrowless}
\newunicodechar{⥸}{\gtrarr}
\newunicodechar{⥹}{\subrarr}
\newunicodechar{⥺}{\leftarrowsubset}
\newunicodechar{⥻}{\suplarr}
\newunicodechar{⥼}{\leftfishtail}
\newunicodechar{⥽}{\rightfishtail}
\newunicodechar{⥾}{\upfishtail}
\newunicodechar{⥿}{\downfishtail}
\newunicodechar{⦀}{\Vvert}
\newunicodechar{⦁}{\mdsmblkcircle}
\newunicodechar{⦂}{\typecolon}
\newunicodechar{⦃}{\lBrace}
\newunicodechar{⦄}{\rBrace}
\newunicodechar{⦅}{\lParen}
\newunicodechar{⦆}{\rParen}
\newunicodechar{⦇}{\llparenthesis}
\newunicodechar{⦈}{\rrparenthesis}
\newunicodechar{⦉}{\llangle}
\newunicodechar{⦊}{\rrangle}
\newunicodechar{⦋}{\lbrackubar}
\newunicodechar{⦌}{\rbrackubar}
\newunicodechar{⦍}{\lbrackultick}
\newunicodechar{⦎}{\rbracklrtick}
\newunicodechar{⦏}{\lbracklltick}
\newunicodechar{⦐}{\rbrackurtick}
\newunicodechar{⦑}{\langledot}
\newunicodechar{⦒}{\rangledot}
\newunicodechar{⦓}{\lparenless}
\newunicodechar{⦔}{\rparengtr}
\newunicodechar{⦕}{\Lparengtr}
\newunicodechar{⦖}{\Rparenless}
\newunicodechar{⦗}{\lblkbrbrak}
\newunicodechar{⦘}{\rblkbrbrak}
\newunicodechar{⦙}{\fourvdots}
\newunicodechar{⦚}{\vzigzag}
\newunicodechar{⦛}{\measuredangleleft}
\newunicodechar{⦜}{\rightanglesqr}
\newunicodechar{⦝}{\rightanglemdot}
\newunicodechar{⦞}{\angles}
\newunicodechar{⦟}{\angdnr}
\newunicodechar{⦠}{\gtlpar}
\newunicodechar{⦡}{\sphericalangleup}
\newunicodechar{⦢}{\turnangle}
\newunicodechar{⦣}{\revangle}
\newunicodechar{⦤}{\angleubar}
\newunicodechar{⦥}{\revangleubar}
\newunicodechar{⦦}{\wideangledown}
\newunicodechar{⦧}{\wideangleup}
\newunicodechar{⦨}{\measanglerutone}
\newunicodechar{⦩}{\measanglelutonw}
\newunicodechar{⦪}{\measanglerdtose}
\newunicodechar{⦫}{\measangleldtosw}
\newunicodechar{⦬}{\measangleurtone}
\newunicodechar{⦭}{\measangleultonw}
\newunicodechar{⦮}{\measangledrtose}
\newunicodechar{⦯}{\measangledltosw}
\newunicodechar{⦰}{\revemptyset}
\newunicodechar{⦱}{\emptysetobar}
\newunicodechar{⦲}{\emptysetocirc}
\newunicodechar{⦳}{\emptysetoarr}
\newunicodechar{⦴}{\emptysetoarrl}
\newunicodechar{⦵}{\circlehbar}
\newunicodechar{⦶}{\circledvert}
\newunicodechar{⦷}{\circledparallel}
\newunicodechar{⦸}{\obslash}
\newunicodechar{⦹}{\operp}
\newunicodechar{⦺}{\obot}
\newunicodechar{⦻}{\olcross}
\newunicodechar{⦼}{\odotslashdot}
\newunicodechar{⦽}{\uparrowoncircle}
\newunicodechar{⦾}{\circledwhitebullet}
\newunicodechar{⦿}{\circledbullet}
\newunicodechar{⧀}{\olessthan}
\newunicodechar{⧁}{\ogreaterthan}
\newunicodechar{⧂}{\cirscir}
\newunicodechar{⧃}{\cirE}
\newunicodechar{⧄}{\boxdiag}
\newunicodechar{⧅}{\boxbslash}
\newunicodechar{⧆}{\boxast}
\newunicodechar{⧇}{\boxcircle}
\newunicodechar{⧈}{\boxbox}
\newunicodechar{⧉}{\boxonbox}
\newunicodechar{⧊}{\triangleodot}
\newunicodechar{⧋}{\triangleubar}
\newunicodechar{⧌}{\triangles}
\newunicodechar{⧍}{\triangleserifs}
\newunicodechar{⧎}{\rtriltri}
\newunicodechar{⧏}{\ltrivb}
\newunicodechar{⧐}{\vbrtri}
\newunicodechar{⧑}{\lfbowtie}
\newunicodechar{⧒}{\rfbowtie}
\newunicodechar{⧓}{\fbowtie}
\newunicodechar{⧔}{\lftimes}
\newunicodechar{⧕}{\rftimes}
\newunicodechar{⧖}{\hourglass}
\newunicodechar{⧗}{\blackhourglass}
\newunicodechar{⧘}{\lvzigzag}
\newunicodechar{⧙}{\rvzigzag}
\newunicodechar{⧚}{\Lvzigzag}
\newunicodechar{⧛}{\Rvzigzag}
\newunicodechar{⧜}{\iinfin}
\newunicodechar{⧝}{\tieinfty}
\newunicodechar{⧞}{\nvinfty}
\newunicodechar{⧟}{\dualmap}
\newunicodechar{⧠}{\laplac}
\newunicodechar{⧡}{\lrtriangleeq}
\newunicodechar{⧢}{\shuffle}
\newunicodechar{⧣}{\eparsl}
\newunicodechar{⧤}{\smeparsl}
\newunicodechar{⧥}{\eqvparsl}
\newunicodechar{⧦}{\gleichstark}
\newunicodechar{⧧}{\thermod}
\newunicodechar{⧨}{\downtriangleleftblack}
\newunicodechar{⧩}{\downtrianglerightblack}
\newunicodechar{⧪}{\blackdiamonddownarrow}
\newunicodechar{⧫}{\mdlgblklozenge}
\newunicodechar{⧬}{\circledownarrow}
\newunicodechar{⧭}{\blackcircledownarrow}
\newunicodechar{⧮}{\errbarsquare}
\newunicodechar{⧯}{\errbarblacksquare}
\newunicodechar{⧰}{\errbardiamond}
\newunicodechar{⧱}{\errbarblackdiamond}
\newunicodechar{⧲}{\errbarcircle}
\newunicodechar{⧳}{\errbarblackcircle}
\newunicodechar{⧴}{\ruledelayed}
\newunicodechar{⧵}{\setminus}
\newunicodechar{⧶}{\dsol}
\newunicodechar{⧷}{\rsolbar}
\newunicodechar{⧸}{\xsol}
\newunicodechar{⧹}{\xbsol}
\newunicodechar{⧺}{\doubleplus}
\newunicodechar{⧻}{\tripleplus}
\newunicodechar{⧼}{\lcurvyangle}
\newunicodechar{⧽}{\rcurvyangle}
\newunicodechar{⧾}{\tplus}
\newunicodechar{⧿}{\tminus}
\newunicodechar{⨀}{\bigodot}
\newunicodechar{⨁}{\bigoplus}
\newunicodechar{⨂}{\bigotimes}
\newunicodechar{⨃}{\bigcupdot}
\newunicodechar{⨄}{\biguplus}
\newunicodechar{⨅}{\bigsqcap}
\newunicodechar{⨆}{\bigsqcup}
\newunicodechar{⨇}{\conjquant}
\newunicodechar{⨈}{\disjquant}
\newunicodechar{⨉}{\bigtimes}
\newunicodechar{⨊}{\modtwosum}
\newunicodechar{⨋}{\sumint}
\newunicodechar{⨌}{\iiiint}
\newunicodechar{⨍}{\intbar}
\newunicodechar{⨎}{\intBar}
\newunicodechar{⨏}{\fint}
\newunicodechar{⨐}{\cirfnint}
\newunicodechar{⨑}{\awint}
\newunicodechar{⨒}{\rppolint}
\newunicodechar{⨓}{\scpolint}
\newunicodechar{⨔}{\npolint}
\newunicodechar{⨕}{\pointint}
\newunicodechar{⨖}{\sqint}
\newunicodechar{⨗}{\intlarhk}
\newunicodechar{⨘}{\intx}
\newunicodechar{⨙}{\intcap}
\newunicodechar{⨚}{\intcup}
\newunicodechar{⨛}{\upint}
\newunicodechar{⨜}{\lowint}
\newunicodechar{⨝}{\Join}
\newunicodechar{⨞}{\bigtriangleleft}
\newunicodechar{⨟}{\zcmp}
\newunicodechar{⨠}{\zpipe}
\newunicodechar{⨡}{\zproject}
\newunicodechar{⨢}{\ringplus}
\newunicodechar{⨣}{\plushat}
\newunicodechar{⨤}{\simplus}
\newunicodechar{⨥}{\plusdot}
\newunicodechar{⨦}{\plussim}
\newunicodechar{⨧}{\plussubtwo}
\newunicodechar{⨨}{\plustrif}
\newunicodechar{⨪}{\minusdot}
\newunicodechar{⨫}{\minusfdots}
\newunicodechar{⨬}{\minusrdots}
\newunicodechar{⨭}{\opluslhrim}
\newunicodechar{⨮}{\oplusrhrim}
\newunicodechar{⨯}{\vectimes}
\newunicodechar{⨰}{\dottimes}
\newunicodechar{⨱}{\timesbar}
\newunicodechar{⨲}{\btimes}
\newunicodechar{⨳}{\smashtimes}
\newunicodechar{⨴}{\otimeslhrim}
\newunicodechar{⨵}{\otimesrhrim}
\newunicodechar{⨶}{\otimeshat}
\newunicodechar{⨷}{\Otimes}
\newunicodechar{⨸}{\odiv}
\newunicodechar{⨹}{\triangleplus}
\newunicodechar{⨺}{\triangleminus}
\newunicodechar{⨻}{\triangletimes}
\newunicodechar{⨼}{\intprod}
\newunicodechar{⨽}{\intprodr}
\newunicodechar{⨾}{\fcmp}
\newunicodechar{⨿}{\amalg}
\newunicodechar{⩀}{\capdot}
\newunicodechar{⩁}{\uminus}
\newunicodechar{⩂}{\barcup}
\newunicodechar{⩃}{\barcap}
\newunicodechar{⩄}{\capwedge}
\newunicodechar{⩅}{\cupvee}
\newunicodechar{⩆}{\cupovercap}
\newunicodechar{⩇}{\capovercup}
\newunicodechar{⩈}{\cupbarcap}
\newunicodechar{⩉}{\capbarcup}
\newunicodechar{⩊}{\twocups}
\newunicodechar{⩋}{\twocaps}
\newunicodechar{⩌}{\closedvarcup}
\newunicodechar{⩍}{\closedvarcap}
\newunicodechar{⩎}{\Sqcap}
\newunicodechar{⩏}{\Sqcup}
\newunicodechar{⩐}{\closedvarcupsmashprod}
\newunicodechar{⩑}{\wedgeodot}
\newunicodechar{⩒}{\veeodot}
\newunicodechar{⩓}{\Wedge}
\newunicodechar{⩔}{\Vee}
\newunicodechar{⩕}{\wedgeonwedge}
\newunicodechar{⩖}{\veeonvee}
\newunicodechar{⩗}{\bigslopedvee}
\newunicodechar{⩘}{\bigslopedwedge}
\newunicodechar{⩙}{\veeonwedge}
\newunicodechar{⩚}{\wedgemidvert}
\newunicodechar{⩛}{\veemidvert}
\newunicodechar{⩜}{\midbarwedge}
\newunicodechar{⩝}{\midbarvee}
\newunicodechar{⩞}{\doublebarwedge}
\newunicodechar{⩟}{\wedgebar}
\newunicodechar{⩠}{\wedgedoublebar}
\newunicodechar{⩡}{\varveebar}
\newunicodechar{⩢}{\doublebarvee}
\newunicodechar{⩣}{\veedoublebar}
\newunicodechar{⩤}{\dsub}
\newunicodechar{⩥}{\rsub}
\newunicodechar{⩦}{\eqdot}
\newunicodechar{⩧}{\dotequiv}
\newunicodechar{⩨}{\equivVert}
\newunicodechar{⩩}{\equivVvert}
\newunicodechar{⩪}{\dotsim}
\newunicodechar{⩫}{\simrdots}
\newunicodechar{⩬}{\simminussim}
\newunicodechar{⩭}{\congdot}
\newunicodechar{⩮}{\asteq}
\newunicodechar{⩯}{\hatapprox}
\newunicodechar{⩰}{\approxeqq}
\newunicodechar{⩱}{\eqqplus}
\newunicodechar{⩲}{\pluseqq}
\newunicodechar{⩳}{\eqqsim}
\newunicodechar{⩴}{\Coloneq}
\newunicodechar{⩵}{\eqeq}
\newunicodechar{⩶}{\eqeqeq}
\newunicodechar{⩷}{\ddotseq}
\newunicodechar{⩸}{\equivDD}
\newunicodechar{⩹}{\ltcir}
\newunicodechar{⩺}{\gtcir}
\newunicodechar{⩻}{\ltquest}
\newunicodechar{⩼}{\gtquest}
\newunicodechar{⩽}{\leqslant}
\newunicodechar{⩾}{\geqslant}
\newunicodechar{⩿}{\lesdot}
\newunicodechar{⪀}{\gesdot}
\newunicodechar{⪁}{\lesdoto}
\newunicodechar{⪂}{\gesdoto}
\newunicodechar{⪃}{\lesdotor}
\newunicodechar{⪄}{\gesdotol}
\newunicodechar{⪅}{\lessapprox}
\newunicodechar{⪆}{\gtrapprox}
\newunicodechar{⪇}{\lneq}
\newunicodechar{⪈}{\gneq}
\newunicodechar{⪉}{\lnapprox}
\newunicodechar{⪊}{\gnapprox}
\newunicodechar{⪋}{\lesseqqgtr}
\newunicodechar{⪌}{\gtreqqless}
\newunicodechar{⪍}{\lsime}
\newunicodechar{⪎}{\gsime}
\newunicodechar{⪏}{\lsimg}
\newunicodechar{⪐}{\gsiml}
\newunicodechar{⪑}{\lgE}
\newunicodechar{⪒}{\glE}
\newunicodechar{⪓}{\lesges}
\newunicodechar{⪔}{\gesles}
\newunicodechar{⪕}{\eqslantless}
\newunicodechar{⪖}{\eqslantgtr}
\newunicodechar{⪗}{\elsdot}
\newunicodechar{⪘}{\egsdot}
\newunicodechar{⪙}{\eqqless}
\newunicodechar{⪚}{\eqqgtr}
\newunicodechar{⪛}{\eqqslantless}
\newunicodechar{⪜}{\eqqslantgtr}
\newunicodechar{⪝}{\simless}
\newunicodechar{⪞}{\simgtr}
\newunicodechar{⪟}{\simlE}
\newunicodechar{⪠}{\simgE}
\newunicodechar{⪡}{\Lt}
\newunicodechar{⪢}{\Gt}
\newunicodechar{⪣}{\partialmeetcontraction}
\newunicodechar{⪤}{\glj}
\newunicodechar{⪥}{\gla}
\newunicodechar{⪦}{\ltcc}
\newunicodechar{⪧}{\gtcc}
\newunicodechar{⪨}{\lescc}
\newunicodechar{⪩}{\gescc}
\newunicodechar{⪪}{\smt}
\newunicodechar{⪫}{\lat}
\newunicodechar{⪬}{\smte}
\newunicodechar{⪭}{\late}
\newunicodechar{⪮}{\bumpeqq}
\newunicodechar{⪯}{\preceq}
\newunicodechar{⪰}{\succeq}
\newunicodechar{⪱}{\precneq}
\newunicodechar{⪲}{\succneq}
\newunicodechar{⪳}{\preceqq}
\newunicodechar{⪴}{\succeqq}
\newunicodechar{⪵}{\precneqq}
\newunicodechar{⪶}{\succneqq}
\newunicodechar{⪷}{\precapprox}
\newunicodechar{⪸}{\succapprox}
\newunicodechar{⪹}{\precnapprox}
\newunicodechar{⪺}{\succnapprox}
\newunicodechar{⪻}{\Prec}
\newunicodechar{⪼}{\Succ}
\newunicodechar{⪽}{\subsetdot}
\newunicodechar{⪾}{\supsetdot}
\newunicodechar{⪿}{\subsetplus}
\newunicodechar{⫀}{\supsetplus}
\newunicodechar{⫁}{\submult}
\newunicodechar{⫂}{\supmult}
\newunicodechar{⫃}{\subedot}
\newunicodechar{⫄}{\supedot}
\newunicodechar{⫅}{\subseteqq}
\newunicodechar{⫆}{\supseteqq}
\newunicodechar{⫇}{\subsim}
\newunicodechar{⫈}{\supsim}
\newunicodechar{⫉}{\subsetapprox}
\newunicodechar{⫊}{\supsetapprox}
\newunicodechar{⫋}{\subsetneqq}
\newunicodechar{⫌}{\supsetneqq}
\newunicodechar{⫍}{\lsqhook}
\newunicodechar{⫎}{\rsqhook}
\newunicodechar{⫏}{\csub}
\newunicodechar{⫐}{\csup}
\newunicodechar{⫑}{\csube}
\newunicodechar{⫒}{\csupe}
\newunicodechar{⫓}{\subsup}
\newunicodechar{⫔}{\supsub}
\newunicodechar{⫕}{\subsub}
\newunicodechar{⫖}{\supsup}
\newunicodechar{⫗}{\suphsub}
\newunicodechar{⫘}{\supdsub}
\newunicodechar{⫙}{\forkv}
\newunicodechar{⫚}{\topfork}
\newunicodechar{⫛}{\mlcp}
\newunicodechar{⫝}{\forksnot}
\newunicodechar{⫞}{\shortlefttack}
\newunicodechar{⫟}{\shortdowntack}
\newunicodechar{⫠}{\shortuptack}
\newunicodechar{⫡}{\perps}
\newunicodechar{⫢}{\vDdash}
\newunicodechar{⫣}{\dashV}
\newunicodechar{⫤}{\Dashv}
\newunicodechar{⫥}{\DashV}
\newunicodechar{⫦}{\varVdash}
\newunicodechar{⫧}{\Barv}
\newunicodechar{⫨}{\vBar}
\newunicodechar{⫩}{\vBarv}
\newunicodechar{⫪}{\barV}
\newunicodechar{⫫}{\Vbar}
\newunicodechar{⫬}{\Not}
\newunicodechar{⫭}{\bNot}
\newunicodechar{⫮}{\revnmid}
\newunicodechar{⫯}{\cirmid}
\newunicodechar{⫰}{\midcir}
\newunicodechar{⫱}{\topcir}
\newunicodechar{⫲}{\nhpar}
\newunicodechar{⫳}{\parsim}
\newunicodechar{⫴}{\interleave}
\newunicodechar{⫵}{\nhVvert}
\newunicodechar{⫶}{\threedotcolon}
\newunicodechar{⫷}{\lllnest}
\newunicodechar{⫸}{\gggnest}
\newunicodechar{⫹}{\leqqslant}
\newunicodechar{⫺}{\geqqslant}
\newunicodechar{⫻}{\trslash}
\newunicodechar{⫼}{\biginterleave}
\newunicodechar{⫽}{\sslash}
\newunicodechar{⫾}{\talloblong}
\newunicodechar{⫿}{\bigtalloblong}
\newunicodechar{⬒}{\squaretopblack}
\newunicodechar{⬓}{\squarebotblack}
\newunicodechar{⬔}{\squareurblack}
\newunicodechar{⬕}{\squarellblack}
\newunicodechar{⬖}{\diamondleftblack}
\newunicodechar{⬗}{\diamondrightblack}
\newunicodechar{⬘}{\diamondtopblack}
\newunicodechar{⬙}{\diamondbotblack}
\newunicodechar{⬚}{\dottedsquare}
\newunicodechar{⬛}{\lgblksquare}
\newunicodechar{⬜}{\lgwhtsquare}
\newunicodechar{⬝}{\vysmblksquare}
\newunicodechar{⬞}{\vysmwhtsquare}
\newunicodechar{⬟}{\pentagonblack}
\newunicodechar{⬠}{\pentagon}
\newunicodechar{⬡}{\varhexagon}
\newunicodechar{⬢}{\varhexagonblack}
\newunicodechar{⬣}{\hexagonblack}
\newunicodechar{⬤}{\lgblkcircle}
\newunicodechar{⬥}{\mdblkdiamond}
\newunicodechar{⬦}{\mdwhtdiamond}
\newunicodechar{⬧}{\mdblklozenge}
\newunicodechar{⬨}{\mdwhtlozenge}
\newunicodechar{⬩}{\smblkdiamond}
\newunicodechar{⬪}{\smblklozenge}
\newunicodechar{⬫}{\smwhtlozenge}
\newunicodechar{⬬}{\blkhorzoval}
\newunicodechar{⬭}{\whthorzoval}
\newunicodechar{⬮}{\blkvertoval}
\newunicodechar{⬯}{\whtvertoval}
\newunicodechar{⬰}{\circleonleftarrow}
\newunicodechar{⬱}{\leftthreearrows}
\newunicodechar{⬲}{\leftarrowonoplus}
\newunicodechar{⬳}{\longleftsquigarrow}
\newunicodechar{⬴}{\nvtwoheadleftarrow}
\newunicodechar{⬵}{\nVtwoheadleftarrow}
\newunicodechar{⬶}{\twoheadmapsfrom}
\newunicodechar{⬷}{\twoheadleftdbkarrow}
\newunicodechar{⬸}{\leftdotarrow}
\newunicodechar{⬹}{\nvleftarrowtail}
\newunicodechar{⬺}{\nVleftarrowtail}
\newunicodechar{⬻}{\twoheadleftarrowtail}
\newunicodechar{⬼}{\nvtwoheadleftarrowtail}
\newunicodechar{⬽}{\nVtwoheadleftarrowtail}
\newunicodechar{⬾}{\leftarrowx}
\newunicodechar{⬿}{\leftcurvedarrow}
\newunicodechar{⭀}{\equalleftarrow}
\newunicodechar{⭁}{\bsimilarleftarrow}
\newunicodechar{⭂}{\leftarrowbackapprox}
\newunicodechar{⭃}{\rightarrowgtr}
\newunicodechar{⭄}{\rightarrowsupset}
\newunicodechar{⭅}{\LLeftarrow}
\newunicodechar{⭆}{\RRightarrow}
\newunicodechar{⭇}{\bsimilarrightarrow}
\newunicodechar{⭈}{\rightarrowbackapprox}
\newunicodechar{⭉}{\similarleftarrow}
\newunicodechar{⭊}{\leftarrowapprox}
\newunicodechar{⭋}{\leftarrowbsimilar}
\newunicodechar{⭌}{\rightarrowbsimilar}
\newunicodechar{⭐}{\medwhitestar}
\newunicodechar{⭑}{\medblackstar}
\newunicodechar{⭒}{\smwhitestar}
\newunicodechar{⭓}{\rightpentagonblack}
\newunicodechar{⭔}{\rightpentagon}
\newunicodechar{〒}{\postalmark}
\newunicodechar{〔}{\lbrbrak}
\newunicodechar{〕}{\rbrbrak}
\newunicodechar{〘}{\Lbrbrak}
\newunicodechar{〙}{\Rbrbrak}
\newunicodechar{〰}{\hzigzag}

\usepackage{pdftexcmds} %
\usepackage{fnpct}
\setfnpct{add-punct-marks=!?:;} %

\usepackage{ellipsis}

\usepackage{fancyvrb}

\usepackage[normalem]{ulem} %
\usepackage{pifont} %
\usepackage{xcolor}
\usepackage{fancyhdr}
\usepackage{xspace}

\usepackage{graphicx} %

\usepackage{tikz}\usetikzlibrary{shapes,calc,arrows,fit,positioning,decorations.pathmorphing,snakes,intersections,shapes.geometric,trees,cd}  %

\usepackage[hyphens]{url} %

\usepackage{afterpage}
\usepackage{wrapfig}
\usepackage{alphalph} %

\usepackage{booktabs} %

\usepackage{array}
\newcolumntype{L}[1]{>{\raggedright\let\newline\\\arraybackslash\hspace{0pt}}m{#1}}
\newcolumntype{C}[1]{>{\centering\let\newline\\\arraybackslash\hspace{0pt}}m{#1}}
\newcolumntype{R}[1]{>{\raggedleft\let\newline\\\arraybackslash\hspace{0pt}}m{#1}}

\usepackage{tabularx} %

\usepackage{environ}

\NewEnviron{tabx}[1]{\renewcommand{\arraystretch}{1.2} %
\tabularx{\textwidth}{@{}#1@{}}
\toprule
\BODY
\bottomrule
}[\endtabularx]

\usepackage{multirow}

 \newcommand{\authorsnote}[2]{}

\usepackage{amsmath}
\usepackage{amsfonts}
\usepackage{amssymb}
\usepackage{mathtools} %

\usepackage{amsthm}

\theoremstyle{definition}

\newtheoremstyle{taskstyle}%
  {.32\baselineskip±.15\baselineskip}%
  {.32\baselineskip±.15\baselineskip}%
  {\it}%
  {}%
  {\bf}%
  {. }%
  { }%
  {}%
\theoremstyle{taskstyle}

\newcounter{subtasknumber}
\newcounter{subtasklabel}
\renewcommand{\thesubtasklabel}{\textbf{\thetasknumber}}

\usepackage[T1]{fontenc}

\usepackage{stmaryrd}

\newcommand{\Expect}{{\rm I\kern-.3em E}}

\newcommand{\D}{\mathcal{D}}

\usepackage{eurosym}

\renewcommand\P{p}
\newcommand\otool{\textsf{\small FPInspector}}
\newcommand\tool{\textsf{\small PETInspector}}
\newcommand\fs{\textsf{\small FPServer}}
\newcommand\cs{\textsf{\small ClientSim}}
\renewcommand{\ae}{\textsf{\small AnaEng}}
\newcommand\tor{Tor BB}
\newcommand\br{Brave}
\newcommand\bi{Blend In}
\newcommand\bl{Blender}
\newcommand\cb{CanvasBlocker}
\newcommand\cfb{CanvasFingerprintBlock}
\newcommand\ca{\cb} %
\newcommand\gl{Glove}
\newcommand\hmf{HideMyFootprint}%
\newcommand\pe{Privacy Extension}
\renewcommand\sf{Stop Fingerprinting}
\renewcommand\to{TotalSpoof}
\newcommand\tp{Tracking Protection}
\newcommand\tr{Trace}
\newcommand\nee{No Enum.\@ Extensions}

\newcommand\ghf{$\text{Ghostery}_\textsc{F}$}
\newcommand\ghc{$\text{Ghostery}_\textsc{C}$}
\newcommand\apf{$\text{Adblock Plus}_\textsc{F}$}
\newcommand\apc{$\text{Adblock Plus}_\textsc{C}$}
\newcommand\uoc{$\text{uBlock Origin}_\textsc{C}$}
\newcommand\uof{$\text{uBlock Origin}_\textsc{F}$}
\newcommand\cdc{$\text{Canvas Defender}_\textsc{C}$}
\newcommand\cdf{$\text{Canvas Defender}_\textsc{F}$}
\newcommand\pbc{$\text{Privacy Badger}_\textsc{C}$}
\newcommand\pbf{$\text{Privacy Badger}_\textsc{F}$}
\newcommand\dc{$\text{Disconnect}_\textsc{C}$}
\newcommand\df{$\text{Disconnect}_\textsc{F}$}

\newcommand\Stor{Tor}
\newcommand\Sbr{\textsc{br}}
\newcommand\Sbi{\textsc{bi}}
\newcommand\Sbl{\textsc{bl}}

\newcommand\Scfb{\textsc{cfb}}
\newcommand\Sca{\textsc{cb}} %
\newcommand\Sgl{\textsc{gl}}
\newcommand\Shmf{\textsc{hmf}}
\newcommand\Spe{\textsc{pe}}
\newcommand\Ssf{\textsc{sf}}
\newcommand\Sto{\textsc{to}}
\newcommand\Stp{\textsc{tp}}
\newcommand\Str{\textsc{tr}}
\newcommand\Snee{\textsc{ne}}

\newcommand\Sghf{$\textsc{gh}_\textsc{F}$}
\newcommand\Sghc{$\textsc{gh}_\textsc{C}$}
\newcommand\Sapf{$\textsc{ap}_\textsc{F}$}
\newcommand\Sapc{$\textsc{ap}_\textsc{C}$}
\newcommand\Suoc{$\textsc{uo}_\textsc{C}$}
\newcommand\Suof{$\textsc{uo}_\textsc{F}$}
\newcommand\Scdc{$\textsc{cd}_\textsc{C}$}
\newcommand\Scdf{$\textsc{cd}_\textsc{F}$}
\newcommand\Spbc{$\textsc{pb}_\textsc{C}$}
\newcommand\Spbf{$\textsc{pb}_\textsc{F}$}
\newcommand\Sdc{$\textsc{d}_\textsc{C}$}
\newcommand\Sdf{$\textsc{d}_\textsc{F}$}

\newcommand\ua{$\mathtt{User{\mhyphen}Agent}$}
\newcommand\al{$\mathtt{Accept{\mhyphen}Language}$}
\mathchardef\mhyphen="2D

\newcommand{\inconc}{\:\cdot}

\newcommand{\squaredot}{\boxdot}
\newcommand{\squarecheck}{\boxplus}
\newcommand{\squaretimes}{\textcolor{red}{\boxtimes}}
\newcommand{\masked}{+}
\newcommand{\yes}{\checkmark}

\newcommand\func{\mathsf{f}}
\newcommand\eff{\mathsf{eff}}

\newcommand\entropy{\mathsf{ent}}

\newcommand\pctbl{\mathsf{prop\_less10}}
\newcommand\pctal{\mathsf{prop\_less1}}

\newcommand\popularity{\mathsf{\# users}}

\newtheorem{preexperiment}{Experiment}

\usepackage{tikz}
\usetikzlibrary{cd}

\newcommand{\spacecut}[1]{}

\begin{document}
\title{The Effectiveness of Privacy Enhancing Technologies\\ against Fingerprinting}

\author{Amit Datta\thanks{The majority of this author's contributions were made while at Carnegie Mellon University.}\\ {\normalsize Snap Inc.} \and Jianan Lu\\ {\normalsize University of California, Berkeley} \and Michael Carl Tschantz\\ {\normalsize International Computer Science Institute}}

\maketitle

\begin{abstract}
We measure how effective Privacy Enhancing Technologies (PETs) are at protecting users from website fingerprinting. Our measurements use both experimental and observational methods.  Experimental methods allow control, precision, and use on new PETs that currently lack a user base. Observational methods enable scale and drawing from the browsers currently in real-world use. By applying experimentally created models of a PET's behavior to an observational data set, our novel hybrid method offers the best of both worlds.  We find the Tor Browser Bundle to be the most effective PET amongst the set we tested.  We find that some PETs have inconsistent behaviors, which can do more harm than good.
\end{abstract}

\section{Introduction}
\label{sec:intro}

Online data aggregators track consumer activities on the Internet to build behavioral profiles. 
Traditional forms of tracking use stateful mechanisms, where the tracker 
(e.g., Google's DoubleClick) places 
an identifier (e.g., an HTTP cookie) with a unique value on the consumer's 
browser or computer. When the consumer visits webpages where the same tracker 
has a presence, their browser automatically sends the identifier value to the tracker. 
This allows the tracker to link these visits to the same consumer. 
Two properties make an identifier good for tracking purposes:
\emph{uniqueness} and \emph{predictability}%
.\footnote{Prior work has used the term stability instead of 
predictability~\cite{eckersley2010unique, nikiforakis2015privaricator, 
laperdrix2017fprandom}. We see stability as a form of predictability where identifier 
values remain identical.} Uniqueness requires that the identifier values are
sufficiently unique among consumers, 
whereas predictability requires the identifier values are predictable for a 
consumer across webpage visits. 

Increased awareness about stateful tracking 
mechanisms has led consumers to take precautions
against them (e.g., by blocking or clearing cookies).
This has spurred the growth of stateless tracking mechanisms, also known as browser
fingerprinting.
A stateless tracker extracts fingerprints from consumers 
as a collection of several attributes of the
browser, operating system, and hardware, typically accessed through Javascript APIs.
Fingerprints collected on websites like \url{panopticlick.eff.org} and \url{amiunique.org/fp} 
demonstrate that they are sufficiently unique and predictable for 
tracking purposes~\cite{eckersley2010unique, laperdrix2016beauty}. 
The list of attributes that can be used 
in fingerprints is rapidly increasing~\cite{mowery2012pixel, acar2013fpdetective, 
acar2014web, fifield2015fingerprinting, englehardt2016online, cao2017cross, 
starov2017xhound}. Studies have also uncovered fingerprinting code on popular 
webpages~\cite{acar2013fpdetective, acar2014web, englehardt2016online}. 

Anti-Fingerprinting Privacy Enhancing Technologies 
(AFPETs) aim to protect consumers against 
fingerprinting by masking, or spoofing, the values of attributes.
For each attribute, they can either (1) \emph{standardize} it, so that all 
of the AFPET's users share the same or one of a small set of attribute values, thereby decreasing the 
uniqueness of fingerprints, or (2) \emph{vary}\footnote{Tor 
developers use the term \emph{randomization}~\cite{torprivacy}. 
We see randomization as an instance of
variation.} it, so that the fingerprints of all the AFPET's 
users vary across webpage visits, thereby decreasing the predictability of fingerprints.

Our goal is to find attributes that AFPETs are not masking (with either
standardization or variation) and to quantify their effects on privacy.
We develop a method
that compares the trackability (i.e., uniqueness and predictability)
of AFPET-modified fingerprints with those of the original fingerprints.
Depending on the goals, AFPET evaluation could depend on the context in 
which the AFPET is used, accounting for features of other users and non-users, 
or be a more theoretical assessment of the AFPET's potential, untied to the vagaries of today.
For example, if the goal of evaluation is to determine which AFPET to use today, one 
would want to know how many other users of the AFPET there are 
since they will form the anonymity set -- the group of other users one will blend in with.
If instead the goal is to determine which AFPET to fund for further development, 
the user numbers of today may matter less than the technical or theoretical
capabilities of the AFPET. Given that no one AFPET evaluation can match all goals,
 we will explore points in the space of possible evaluations while focusing more on
prospective evaluations.
\subsection{Methods}
\label{sec:intro-methods}

First, we consider a more theoretical, experimental analysis that 
directly looks at an AFPET's ability to mask attributes. %
This method runs browsers with and without an AFPET installed to determine 
which attributes the AFPET masks, either by standardizing or varying its value. 
For this purpose, we develop an experimental framework, \tool, which has 
three components: the \emph{fingerprinting server} (\fs), which collects fingerprints from visitors,  
the \emph{client simulator} (\cs), which simulates 
consumers and drives them to \fs\ with and without AFPETs, 
and the \emph{analysis engine} (\ae), which compares fingerprints 
across clients to produce a \emph{mask model} characterizing AFPET behaviors.
This tool can be applied to new AFPETs that currently lack a user base.
This experimental method does not require access to the source code of AFPETs.
However, it does not tell us which attributes are the most important to mask.

Next, we consider a highly context-dependent, observational method.
Websites like \url{panopticlick.eff.org} and 
\url{amiunique.org/fp} obtain large sets of real-world fingerprints, 
revealing which are the most trackable (i.e., unique and predictable).
In principle, these datasets can be studied to evaluate an 
AFPET by selecting the fingerprints generated by users of that AFPET and, for each such 
fingerprint, checking how trackable they are compared to other fingerprints in the dataset.
We have implemented the core task of measuring trackablity as a tool,
\otool{}, which simulates a simple tracker and computes statistics
quantifying anonymity, such as entropy.
In practice, however, such observational datasets may contain too few users of an AFPET, 
especially for new ones, for \otool{} evaluate it.
Furthermore, in some cases, it may be difficult to determine which fingerprints 
correspond to which AFPETs.
Thus, utilizing such a dataset requires a more nuanced approach.

\begin{figure*}
\centering
\includegraphics[width=0.95\linewidth]{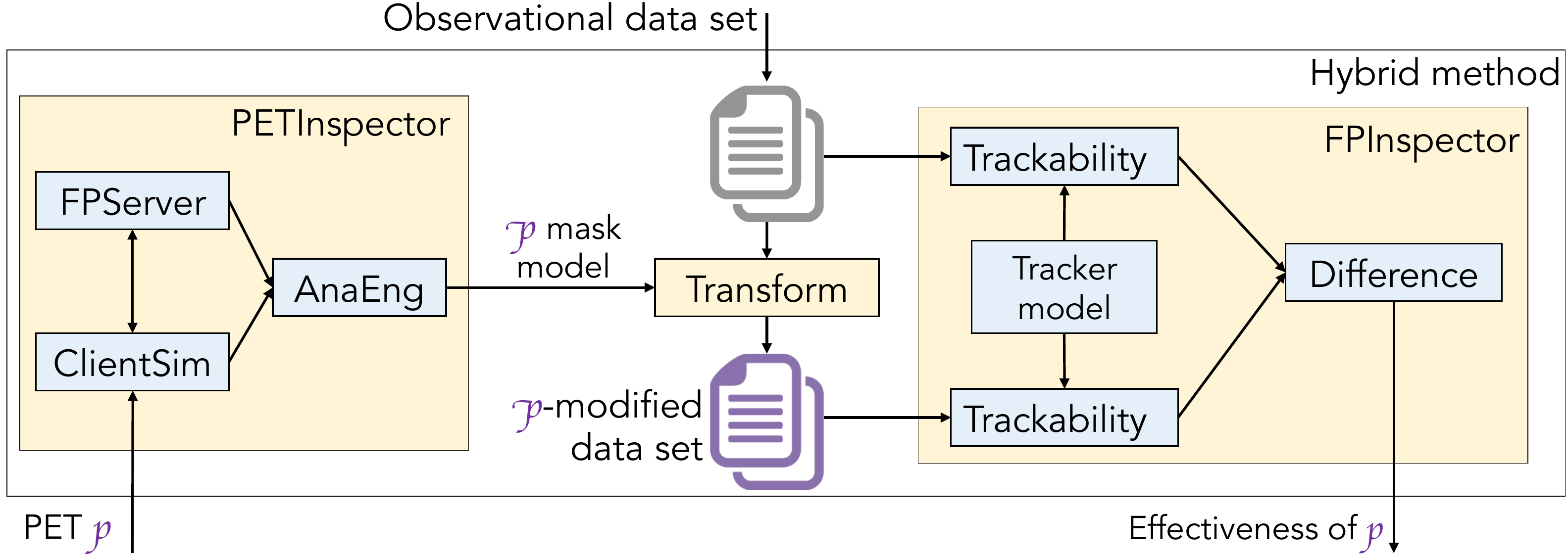}
\caption{Hybrid method for AFPET evaluation}
\label{fig:method}
\end{figure*}

Then, we develop a novel hybrid method combining observational and
experimental data to enable the evaluation of AFPETs with low or no
usage within the context of the browsers used today but without access
to the AFPET source code.
Our hybrid method combines \otool{} with \tool{} as 
outlined in Figure~\ref{fig:method}.
It contextualizes the mask model produced by \tool{}
by applying it to an observational dataset of
real-world fingerprints to produce a counterfactual dataset
representing what the browsers would look like to trackers had everyone used the AFPET.
By comparing the trackabilities on the two datasets, we can evaluate the effectiveness 
of the AFPET.
By parametrically leveraging data from ongoing,
large-scale measurement studies, our results may be updated for the
ever changing landscape of browsers with little additional work.

Finally, we adjust the hybrid method to take into account the number
of users an AFPET has.
This shifts the analysis even further in the direction of examining
the PET's current abilities over its theoretical possibilities.

\subsection{Results}
\label{sec:intro-results}
Using \tool, we resolved with high confidence whether
$15$ AFPETs explicitly claiming to 
protect against fingerprinting 
mask $20$ attributes of Firefox and $18$ attributes of Chrome.
We also looked at $11$ other popular \emph{blacklisting PETs}
(BLPETs), which operate by blacklisting domains known to engage in
tracking.
While they do not make a claim of protecting against fingerprinting,
they should not make matters worse by giving browsers a more unique
fingerprint, a property we check them for.

We found that all but the Tor Browser Bundle masked $9$ or fewer of
the resolved attributes, at least in their default configurations.
In particular, we found that Tor left a single 
attribute, $\mathtt{platform}$, unmasked
while all others left at least $12$ attributes unmasked.
\tool{} also uncovered undocumented behaviors and 
inconsistencies in how some PETs modify various attributes:
\begin{itemize}

\item Brave Browser spoofs the \ua\ to appear like Chrome. However, it modifies the 
\al\ header, $\mathtt{language}$ and $\mathtt{plugins}$ differently than baseline Chrome. 
To a tracker, this can make \br\ users stand out from other Chrome users. We have raised
the issue with Brave developers
and have received comments from them
acknowledging the issue~\cite{tadatitam18muon,tadatitam18laptop}.

\item 
Privacy Badger and Firefox's tracking protection configuration implement Do Not Track
differently. While both send the $\mathtt{Dnt}$ header, only Firefox sets the
$\mathtt{doNotTrack}$ variable in JavaScript's navigator object. 
As a result, web-services 
which only use JavaScript to detect the Do Not Track choice will not be able to do so
for Privacy Badger users. Furthermore, this inconsistency may make Privacy Badger's
users stand out, making them easier to track.
We raised this issue with Privacy Badger 
developers who have since fixed the 
issue~\cite{tadatitam18privacy-badger}.

\item HideMyFootprint randomizes the \ua\ header, 
while not modifying the $\mathtt{platform}$. This leads to inconsistencies like the 
\ua\ containing \emph{Windows NT 10.0} on a \emph{Linux x86\_64} $\mathtt{platform}$. 
Moreover it sends an additional $\mathtt{Pragma}$ header, which can make %
users distinguishable.

\item While the Tor Browser is able to conceal the operating system by spoofing 
attributes like the \ua\ and $\mathtt{cpu\ class}$, the $\mathtt{javascript\ 
fonts}$ revealed by different browsing platforms can reveal that information. 
\end{itemize}

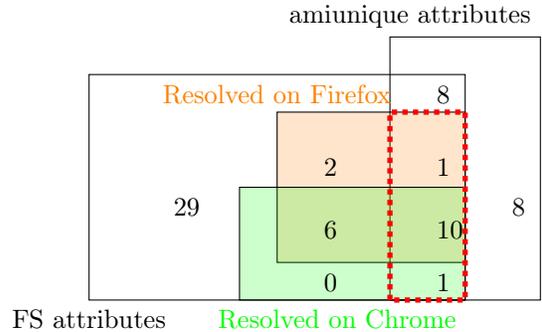
\begin{figure}
\centering
\begin{tikzpicture}
\draw (0.0,0.0) rectangle +(5,3);
\draw (4,0) rectangle +(2,3.5);
\draw[fill=orange, fill opacity=0.2] (2.5,0.5) rectangle +(2.5,2);
\draw[fill=green, fill opacity=0.2] (2,0) rectangle +(3,1.5);
\node[below] at (0,0) {FS attributes};
\node[above left] at (6,3.5) {amiunique attributes};
\node[text=orange, above] at (2.5,2.5) {Resolved on Firefox};
\node[below right] at (3,2) {2};	%
\node[below right] at (4.5,2) {1}; %
\node[text=green, below] at (3.3,0) {Resolved on Chrome};
\node[above right] at (3,0) {0}; %
\node[above right] at (4.5,0) {1}; %
\node[above right] at (3,0.7) {6}; %
\node[above right] at (4.5,0.7) {10}; %
\node[above right] at (4.5,2.5) {8}; %
\node[above right] at (5.5,1) {8}; %
\node[above right] at (1,1) {29}; %
\draw[color=red, line width=0.7mm, dotted] (4,0) rectangle +(1,2.5);
\end{tikzpicture} %
\caption{\fs\ collects $49$ attributes, of which $29$ remain unexercised by \cs. Amiunique dataset has $28$ unique attributes, of which $8$ aren't collected by \fs\ and $8$ are collected but remain unexercised by \cs. The red dotted rectangle represents the intersection of attributes exercised by \cs\ and present in the amiunique dataset and used for our hybrid evaluation.}\label{fig:attributesvenn}
\end{figure}

We found $6$ AFPETs which masked $4$ attributes, but they did not all
mask the same set of attributes. To break such ties, we used the hybrid method.
We used a pre-existing dataset of over 25,000 real-world fingerprints 
collected on the website 
\url{amiunique.org}.\footnote{The dataset was graciously provided to 
us by Pierre Laperdrix, one of the creators of \url{amiunique.org}.}
Of the 18--20 attributes resolved for each AFPET, only $12$ appear in the
\texttt{amiunique.org} dataset. 
Figure~\ref{fig:attributesvenn} provides an overview of how we selected attributes for our 
hybrid evaluation. 
For these $12$ attributes, the hybrid method generates a set of 
PET-modified fingerprints from the 
original fingerprints and measures the effectiveness of the $15$ AFPETs with \otool. 

Our hybrid method finds that even with just $12$ attributes, $13$ of the
$15$ AFPETs do not provide much protection over
using no PET at all, decreasing the entropy revealed from about $13$ bits without any PET 
to $11$ bits with the AFPET. 
It finds the Tor Browser Bundle (\tor) to be most effective, 
revealing under $3$ bits of entropy.

Recognizing that automation has its limitations,
we manually analyzed some of the more interesting findings.
We found that some AFPETs performed better when switched out of their
default configuration.
While we find that some do mask attributes labeled as
inconclusive by \tool,
we did not find any falsely labeled as masked or as unmasked.

A source of entropy for \tor\ fingerprints is the revealed screen resolution, which is only partly masked.
\tor\ reveals partial information about the screen resolution of its users using 
a spoofing strategy which depends on the true resolution for usability reasons. 
We explore a space of alternate spoofing strategies and find some to be just as effective according to our metrics despite being more usable, by utilizing more pixels
on average for browsing than Tor. 
\subsection{Interpretation}
\label{sec:intro-interpretation}

BLPETs do not claim to protect against fingerprinting and AFPETs do
not claim to protect against all forms of fingerprint.
Thus, our results should not necessarily be interpreted as finding
flaws in PETs.
An AFPET that masks the one and only one attribute that it claims to
mask behaves as advertised.

Nevertheless, our tool can be useful for AFPET developers.
It can test whether they masked the attributes they intended to do
so and help ensure that their documentation is correct.
Indeed, we found that AFPET Trace and Tor did not mask all of the attributes
that their documentation claimed that they did
(Table~\ref{tab:studybullet}).

BLPETs do not claim to protect against fingerprinting, but
even they should avoid making browsers more
fingerprintable than they already are.
For example, we found that Privacy Badger made fingerprinting easier
by modifying an attribute in a particular and undocumented way.
Despite not making any anti-fingerprinting claims, its
developers took this result seriously and updated Privacy Badger since
it was an unintended side effect of their approach to privacy.

For consumers and their advocates attempting to select a PET, our
results are also useful beyond the pre-existing, and sometimes
flawed, documentation.
In addition to double checking documentation, 
such consumers may be less concerned with whether PETs meet their
specifications than their overall effectiveness
Our hybrid method allows us to rank the PETs in overall terms of how
well they prevent fingerprinting (Table~\ref{tab:summaries}).
This fine-grained information is not offered by the documentation of
any PET.

Our results are best understood as providing a lower bound on how much
room for improvement remains for AFPETs.
Since AFPETs might mask attributes that we do not test,
we cannot claim to have captured all the work that went into developing an AFPET.
Our lower bound on remaining work is sound in that when \tool{} claims that
an AFPET leaves an attribute unmasked,
it really is not masking it, is not varying the attribute often enough to be
effective, or is not masking enough values of the attribute to protect our test browser platforms.
All three possibilities are concerning.

Our bound is only a lower bound since, by resolving only the status of 18--20
attributes of each browsing platform, we might 
label some attributes in need of masking
as inconclusive.
More attributes can be added to our tools, but the set of possible
attributes is open ended and finding platforms that differ in all attributes can be difficult.
In fact, we are already aware of $30$ %
attributes that we can measure but could not make a 
high-confidence masking determination for
due to having insufficient diversity in their values across our experimental browsing 
platforms.
However, given the long list of issues with the attributes
we did test, we may have already found enough to keep AFPET developers busy.

As for our quantification of the importance of attributes,
it is based on the AmIUnique observational data,
which is not perfectly representative.
Given better data, our tool, without modifications, can
produce more accurate measurements, and
any inaccuracy in our quantification will not affect
the qualitative result that an attribute is left unmasked.

\subsection{Contributions}
\label{sec:intro-contributions}

We make the following main contributions:
\begin{itemize}

\item We develop an experimental framework (\tool) to verify how $15$~AFPETs
(and $11$~BLPETs)
mask 18--20 different attributes each.
 By obtaining a more complete picture of 
PETs' behaviors, we uncover some inconsistencies and peculiarities
(Section~\ref{sec:experiment}).

\item We develop a hybrid method for evaluating AFPETs from an observational 
dataset of real-world fingerprints and apply it to evaluate $15$ AFPETs.
We find \tor\ to be the most effective AFPET among the ones we evaluate
(Section~\ref{sec:datadriven}). 

\item We adjust the hybrid method to also consider the current number
  of users each AFPET has (Section~\ref{sec:pop}).

\item We explore a space of alternate spoofing strategies for screen resolution by \tor\ and 
uncover some which have higher screen utilization than \tor, but are just as effective 
(Section~\ref{sec:tradeoff}).
\end{itemize}

\section{Prior Work}

Prior work finds that various attributes are trackable by measuring the
uniqueness and predictability of fingerprints collected from 
real-world browsing 
platforms~\cite{yen2012host, eckersley2010unique, laperdrix2016beauty}. 
However, few studies evaluate the effectiveness of AFPETs against fingerprinting. 

Many prior studies have focused on BLPETs, which use blacklists
to block known tracking domains and scripts. %
Since BLPETs try to prevent the consumer's browser from 
interacting with trackers, metrics suggestive of successful interactions 
(e.g., third-party requests sent, cookies placed, etc.)
are good indicators of BLPET effectiveness. Studies have evaluated BLPETs by 
comparing these metrics between browsers with and without the BLPET when visiting 
popular websites~\cite{ikram2017towards, 
krishnamurthy2006generating, roesner2012detecting, mayer2012third, 
merzdovnik2017block, hill2015ublock, kontaxis2015tracking, englehardt2016online}. 
FPGuard takes a blacklisting strategy to protect against 
fingerprinting: it uses heuristics to identify fingerprinting domains and  
blocks them~\cite{faizkhademi2015fpguard}. 
Guly\'as~et~al.\@ study the tradeoff between a BLPET suppressing some trackers but also leading to the browser having a more unique fingerprint by being a rare browser extension~\cite{gulyas18wpes}.

Most AFPETs protect against fingerprinting by spoofing browser, operating system 
and hardware characteristics, without blocking specific domains and scripts. For example,
PETs like the Tor Browser standardize various 
attribute values~\cite{torprivacy}, whereas others like 
PriVaricator~\cite{nikiforakis2015privaricator}, FP-Block~\cite{torres2015fp}, 
Blink~\cite{laperdrix2015mitigating}, and FPRandom~\cite{laperdrix2017fprandom} vary 
them. Metrics used for evaluating BLPETs would not be able to 
meaningfully evaluate these AFPETs. Some studies have evaluated attribute varying AFPETs 
by observing variations in fingerprints when using these AFPETs (e.g., 
\cite{nikiforakis2015privaricator, laperdrix2017fprandom}). 
Vastel~et~al.\@ look at how AFPETs can introduce inconsistencies between attributes leading to a more unique fingerprint~\cite{vastel18usenix}.
Our work differ from these by using a combination of experimental and observational data to more thoroughly evaluate AFPETs.

\section{Trackers and PETs}\label{sec:notation}

When a user visits a webpage, trackers can have the user's
browser execute code that requests information about 
the user's browsing platform, including their hardware,
operating system, and the browser itself.
The leftmost column of Table~\ref{tab:studybullet} provides a list of
$49$ attributes known to be good candidates for
fingerprinting.\footnote{We do not consider
  more sophisticated cross-device~\cite{zimmeck2017privacy} and
  cross-browser~\cite{cao2017cross} fingerprints, which aim to link
  together different browsing platforms originating from the same
  consumer.}
The tracker can combine multiple attributes $a_1,\ldots,a_n$ to
compute a \emph{fingerprint}
$id(b) = \langle a_1(b),\ldots,a_n(b)\rangle$ of the browsing platform
$b$ where $a_i(b)$ represents the value of attribute $a_i$ for the
platform $b$.
A tracker can use fingerprints to identify browsing
platforms visiting two websites as being the same one.
The more unique the fingerprint is for each user, the fewer false matches
the tracker will produce in linking two different users.
The more predictable (ideally, unchanging) the fingerprint is as a
user goes from website to website, the fewer matches the tracker will
miss.

To protect themselves from fingerprinting, consumers can install AFPETs
on their browsing platform, which can decrease the uniqueness or
predictability of the platform's fingerprints.
Upon installing a PET $\P$, the consumer's 
browsing platform $b$ is modified to $\P(b)$. 
As a result, the tracker now interacts with 
$\P(b)$ and extracts the fingerprint $id(\P(b))$. 

In this study, we look at three types of PETs:
\let\theenumitemp=\theenumi
\renewcommand{\theenumi}{\Roman{enumi}}
\begin{enumerate}%

\item\label{cat:stan}\textbf{Attribute standardizing.} 
These AFPETs reveal one (full standardization) or one of a 
small set of possible values (partial standardization) for an attribute. 
Full standardization makes all AFPET users appear identical, whereas
partial standardization makes them appear to be from a small number of groups, 
with respect to that attribute. An AFPET may choose partial over full standardization 
if spoofing the attribute value has usability implications. 

\item\label{cat:vary} \textbf{Attribute varying.} 
These AFPETs vary the value of an attribute so that the values 
of each user varies across browsing activities. Such variations may 
affect both the predictability and the uniqueness of the revealed attribute. 
Laperdrix et al.~\cite{laperdrix2017fprandom} show that variation AFPETs can 
vary attributes in a manner that reduces their usability impact.

\item\label{cat:block}
\textbf{Interaction blocking.} 
These BLPETs block some or all interactions between the browsing platform and trackers. 
They rely on a blacklist (e.g., EasyPrivacy) to block interactions matching known 
tracking patterns. %
Trackers interacting with browsing platforms with these PETs 
receive an error message instead of the true fingerprints. 

\end{enumerate}
\renewcommand{\theenumi}{\theenumitemp}

We are primarily interested in evaluating and
comparing AFPETs %
that modify the attribute values either by
standardizing (\ref{cat:stan}) or varying (\ref{cat:vary}) their values.
In some places, we comment on BLPETs that block interactions with known
trackers (\ref{cat:block}).
We do so even for BLPETs not claiming to be AFPETs since they are popular, have been the
subject of past evaluation studies, %
have the potential to unintentionally make fingerprints more unique (as we find with Privacy Badger),
and can be used as AFPETs.
(None of BLPETs that we test suggests using it as an AFPET. 
The BLPET FPGuard did~\cite{faizkhademi2015fpguard},
but we could not find it publicly available for testing.)
However, we do not directly compare them to the AFPETs since 
they do not purport to modify any attributes explicitly, and their
quality depends upon the quality of their blacklists,
necessitating a different form of evaluation.

We leave out of scope PETs that protect against
fingerprinting by blocking scripts (e.g., NoScript~\cite{noscript} and
ScriptSafe~\cite{scriptsafe}) since they %
have considerable impact on usability~\cite{ikram2017towards}.
We also leave out PETs
like Noiszy (\url{noiszy.com}), Internet Noise (\url{makeinternetnoise.com}),  
and AdNauseum (\url{adnauseam.io}) %
that do not attempt to prevent tracking but rather to make it
pointless by injecting noise into the user's history with fake clicks
and website visits.

In this paper, we consider a total of $26$ PETs. 
We assign each PET a unique abbreviation, which we use in some tables. 
When the distinction is needed, we add either a ``\textsc{c}'' for
Chrome or an ``\textsc{f}'' for Firefox to the name or abbreviation of
PETs with versions for both browsers.
We present the full list of PETs, their abbreviations, baseline browser, and strategy
in Table~\ref{tab:pets}.  $23$ of the $26$ PETs are
extensions for Chrome and Firefox\footnote{Extensions for Firefox are called add-ons.} 
(the two most popular desktop browsers at the time
of writing), two are full browsers, and one is a browser configuration. 
Among browser extensions, $11$ are for Chrome, and $12$ are for 
Firefox.\footnote{Several Firefox extensions have been rendered incompatible with 
Firefox $57.0+$ due to a transition in their add-on policies.} $12$ of 
the $23$ extensions are pairs of $6$ extensions available for both Chrome and Firefox. 
$15$ of the $26$ PETs are AFPETs and purport 
to either standardize or vary attribute values, while $11$ others are popular BLPETs.
Some PETs assume mixed strategies. For example, 
\br, \hmf, and \tr\ modify some attributes in addition to 
blocking some types of interactions. The distribution of the strategies adopted by 
these $26$ PETs are presented in Figure~\ref{fig:strategyvenn}.

\begin{table}
\centering
\setlength\tabcolsep{5pt}
\caption{List of PETs we study, their abbreviation, and strategy to protection. 
Most PETs are browser extensions, * indicates full browsers, 
and ** indicates browser configurations.
}\label{tab:pets}
\small
\renewcommand{\arraystretch}{1.3} %
\begin{tabular}{@{}llll@{}}%
\toprule
PET & Abbr. & Strategy & \begin{tabular}{@{}l@{}}AFPET\\ claim?\end{tabular}\\
\midrule
\multicolumn{4}{c}{Chrome PETs} \\
\midrule

CanvasFingerprintBlock~\cite{canvasfingerprintblock}  &  \Scfb    & \ref{cat:stan} & \yes \\ %

Privacy Extension~\cite{privacyextension}  &  \Spe    & \ref{cat:stan}  & \yes \\ %

Brave~\cite{bravebrowser}  &  \Sbr *    & \ref{cat:stan}+\ref{cat:block} & \yes \\ %

Canvas Defender~\cite{multiloginapp}  &  \Scdc    & \ref{cat:vary} & \yes\\ %

Glove~\cite{glove}  &  \Sgl    & \ref{cat:vary}  & \yes\\ %

HideMyFootprint~\cite{hidemyfootprint}  &  \Shmf    & \ref{cat:vary}+\ref{cat:block}& \yes \\ %

Trace~\cite{trace} &  \Str     & \ref{cat:vary}+\ref{cat:block}  & \yes \\ 

Adblock Plus~\cite{adblockplus}  &  \Sapc    &  \ref{cat:block}  \\ %

Disconnect~\cite{disconnect}  &  \Sdc    & \ref{cat:block}   \\ %

Ghostery~\cite{ghostery}  &  \Sghc    & \ref{cat:block}   \\ %

Privacy Badger~\cite{privacybadger}  &  \Spbc  & \ref{cat:block} \\

uBlock Origin~\cite{ublockorigin}   &  \Suoc     & \ref{cat:block}   \\ %

\midrule
\multicolumn{4}{c}{Firefox PETs} \\
\midrule

Blend In~\cite{blendin}  &  \Sbi    & \ref{cat:stan} & \yes \\ %

Blender~\cite{blender}  &  \Sbl    &  \ref{cat:stan} & \yes\\ %

No Enum. Extensions~\cite{nee}  &  \Snee  & \ref{cat:stan}  & \yes\\

Stop Fingerprinting~\cite{stopfingerprinting}  &  \Ssf    & \ref{cat:stan}  & \yes\\ %

Tor Browser Bundle~\cite{torprivacy}  &  \Stor  *    & \ref{cat:stan}  & \yes\\ %

TotalSpoof~\cite{totalspoof}  &  \Sto    & \ref{cat:stan}  & \yes \\ %

Canvas Defender~\cite{multiloginapp}  &  \Scdf    & \ref{cat:vary}  & \yes\\ %

CanvasBlocker~\cite{canvasblocker}  &  \Sca    & \ref{cat:vary} & \yes \\ %
Adblock Plus~\cite{adblockplus}  &  \Sapf     & \ref{cat:block}   \\ %

Disconnect~\cite{disconnect}  &  \Sdf    & \ref{cat:block}   \\ %

Ghostery~\cite{ghostery}    &  \Sghf    & \ref{cat:block}   \\ %

Privacy Badger~\cite{privacybadger}  &  \Spbf  & \ref{cat:block} \\

Tracking Protection~\cite{trackingprotection}  &  \Stp  **   & \ref{cat:block}   \\ %

uBlock Origin~\cite{ublockorigin} &  \Suof      & \ref{cat:block}   \\ %

\bottomrule
\end{tabular}

\end{table}

\def\firstcircle{(0,0) circle (1.5cm)} 
\def\secondcircle{(60:2cm) circle (1.5cm)} 
\def\thirdcircle{(0:2cm) circle (1.5cm)} 
\begin{figure}
\centering
\begin{tikzpicture} [scale=0.8]
\begin{scope}
\draw \firstcircle node[below,shift={(-1.5cm,-1.5cm)}] {Attribute Standardizing} node[below left] {$8$} node[above,shift={(0.2cm,0.7cm)}] {$1$}; 
\draw \secondcircle node [above,shift={(0cm,1.5cm)}] {Interaction Blocking} node[above] {$11$} node[below,shift={(0cm,-0.7cm)}] {$0$} node[below,shift={(0cm,-1.5cm)}] {$0$}; 
\draw \thirdcircle node [below,shift={(1.5cm,-1.5cm)}] {Attribute Varying} node[below right] {$4$}  node[above,shift={(-0.2cm,0.7cm)}] {$2$};
\end{scope} 
\end{tikzpicture}
\caption{Strategies of the $26$ PETs we evaluate}\label{fig:strategyvenn}
\end{figure}

We went over the documentation of the PETs to uncover how they purport to 
modify attributes. For all PETs that explicitly document masking an attribute, 
we place a $\square$ in the corresponding cell in Table~\ref{tab:studybullet}. 
This approach is similar to how Torres et al.\ produce their comparison 
table~\cite[Table 1]{torres2015fp}. However, the documentation is not always clear 
about which attributes are masked. 
One can obtain additional clarity from the programs themselves for open-sourced PETs, 
but source-code analyses
cannot be applied to proprietary PETs. As a result, the $\square$s in 
Table~\ref{tab:studybullet} may not reflect the full picture of how PETs mask
attributes. 
Next, we demonstrate how we use our experimental
method to obtain a more complete picture of the masking behavior. 

\section{Experimental Evaluation of AFPETs}
\label{sec:experiment}

We now consider an experimental, or test-based, approach to AFPET evaluation conducted with artificial users.
These artificial users browse on platforms differing in
whether they have an AFPET installed. 
By comparing fingerprints generated by these artificial users, 
we infer which attributes the AFPET is masking.
We use the degree of masking by each AFPET as an evaluation metric.

Below, we discuss this method and our experimental framework, \tool{},
implementing it.
We then describe an experiment we ran using \tool{} and the results.
The results show that while one could instead look to an AFPET's
documentation for information on which attributes it masks, the
documentation sometimes provides an incomplete picture of an AFPET's
behavior.
We end with a discussion of this method's limitations.

\subsection{Method}

Our experimental framework, \tool{}, is composed of three parts.
The \emph{client simulator}, \cs, creates and drives
experimental browsing platforms, with and without various AFPETs
installed, to visit a server.
The \emph{fingerprinting server}, \fs, collects fingerprints
when the browsing platforms, driven by \cs, visit it.
The \emph{analysis engine}, \ae, compares
fingerprints across clients to 
detect whether an AFPET varies, standardizes, or does not mask the value of an attribute. 
To observe these behaviors, \ae{} compares the value of the attribute on the browsing 
platform without any AFPET (i.e., on the baseline browser)
with the value when an AFPET is installed.

\fs{} plays the role of an online tracker with
the browsers and \fs{} interacting to simulate fingerprinting in the wild.
The components surrounding this simulation produce a
view of AFPETs' effects on fingerprints, 
with \cs{} telling \ae{} which fingerprints correspond
to which AFPETs.

\paragraph{Client Simulator}%

\cs\ drives simulated clients using browsing platforms with different
configurations to visit \fs.
For each base configuration and PET, \cs\ simulates a pair of clients only differing on 
whether the PET is installed, to allow the isolation of the PET's effects.

We choose the base configurations to exercise a wide range of
attribute values in hopes of triggering an AFPET's masking behavior even
when the masking is partial.
For attributes that differ across the browsing platforms, 
we can detect whether an AFPET was standardizing them by comparing their values across
the platforms. Thus, we set up \cs\ to exercise control over as many attributes as possible. 
\cs\ simulates browsing platforms either locally on a computer or on 
pre-configured VirtualBox virtual machines~\cite{watson2008virtualbox} to exercise 
control over many of these attributes. 

\cs\ sets up virtual machines and configures them according to stated preferences, including 
simulating different fonts, 
timezones, languages, and screen properties. 
\cs\ installs different fonts by adding a TrueType font file (\texttt{.ttf} file) to the \texttt{.fonts} 
folder. Both Firefox and Chrome allow fonts from this folder to be rendered on a webpage.
To set timezones, \cs\ uses the \texttt{timedatectl} command available by default on Linux.
\cs\ specifies the language using the \texttt{locale-gen} and by changing the \texttt{LANG}
environment variable. 
Moreover, \cs\ installs the corresponding Firefox language pack. 
Chrome does not have different installers for different languages, instead switching language
based on the \texttt{LANG} environment variable. For screen attributes, specifically 
$\mathtt{Height}$, $\mathtt{Width}$ and $\mathtt{Depth}$, \cs\ uses the display server
\texttt{Xvfb}. 
For native browsing platforms including Mac, Linux and Windows, we used configurations in which we found them.

Exercising control over some attributes is difficult. Some attributes require 
modifications to hardware (e.g., $\mathtt{max\ touch\ points}$) or operating system libraries
(e.g., $\mathtt{math}$ attributes). Screen attributes other than $\mathtt{Height}$, $
\mathtt{Width}$ and $\mathtt{Depth}$ cannot be simulated using \texttt{Xvfb}. 
Moreover, we restrict \cs\ to configuring attributes in the operating system 
while leaving browser settings intact. We do this to prevent 
re-configuring every browser instance after launch which may nullify
the effect of the installed AFPET.
As a result, \cs\ does not exercise
$\mathtt{openDB}$, $\mathtt{indexedDB}$, two $\mathtt{storage}$ attributes, and
six $\mathtt{header}$ attributes.
\cs\ does not configure $\mathtt{plugins}$ since most plugins no longer work 
on Firefox~\cite{firefox-plugins} or Chrome~\cite{chrome-plugins} and 
they are gradually being phased out. 
We do not exercise the $\mathtt{DNT\ enabled}$ 
attribute since it conflicts with \tp.
Nor do we exercise the $\mathtt{adBlock\ installed}$
(a heuristic Javascript test that attempts to insert an ad script
into the page)
and $\mathtt{has\ lied\ with}$ attributes (which checks whether the browser lied about
certain attributes) since they are aimed at detecting various PET behaviors.

After setting up a simulated browsing platform, \cs\ drives browser instances on them
using the Selenium Webdriver~\cite{salunke2014selenium} to \fs. \cs\ launches
a browser in its original configuration or with a PET installed. 
For PETs that are browser extensions, 
\cs\ utilizes Selenium's \texttt{add\_extension} feature on the PETs' 
\texttt{.crx} (Chrome) or \texttt{.xpi} (Firefox) extension files to launch a
PET-enabled browser instance.
For PETs that configure browsers,
\cs\ launches browser instances with specific settings. 
For PETs that are full browsers, \cs{} uses binaries (for \br) or specialized software 
(tbselenium~\cite{acar18tbselenium-github,acar18tbselenium-pypi} for \tor). 

The browser instances interact with \fs\ in a specified pattern of reloads and 
idling to provide insights about the modification behavior of PETs.
In hopes of triggering a PET's ability to mask by varying attribute values,
\cs{} drives its browsers across various boundaries that may cause the PET 
to refresh its spoofed value:
\emph{reloads} of a single domain, visits to different \emph{domains} 
(we give \fs{} two domain names), and browsing across \emph{sessions}.
We define a session to 
browsing separated by 45 minutes of down time,
following Mozilla's definition 
of a session as a continuous period of user activity in the browser, 
where successive events are separated by no more than 30 
minutes~\cite{ulmer10blog}.

\paragraph{Fingerprinting Server}%

\fs{} collects attributes known to be helpful for fingerprinting. 
Specifically, we set up \fs\ to collect attributes collected by the open-source fingerprinting programs
used by FPCentral~\cite{fpcentral} and Panopticlick~\cite{panopticlick}, often by re-using their code. %
We list these attributes in the first column of Table~\ref{tab:studybullet}.
Similar to websites like \url{panopticlick.eff.org} and \url{amiunique.org/fp}, any browser
visiting these domains can view their fingerprint, while a copy is stored on the server. 

\fs\ has minor modifications in the attributes it collects and how it collects them. 
For example, \fs\ detects additional \textit{Noto} fonts, which ship by default on \tor. 
Moreover, \fs\ does not place cookies on 
the browsers visiting our domains, which FPCentral and Panopticlick use to identify
returning visitors.

\begin{figure*}
\centering
\includegraphics[width=0.8\linewidth]{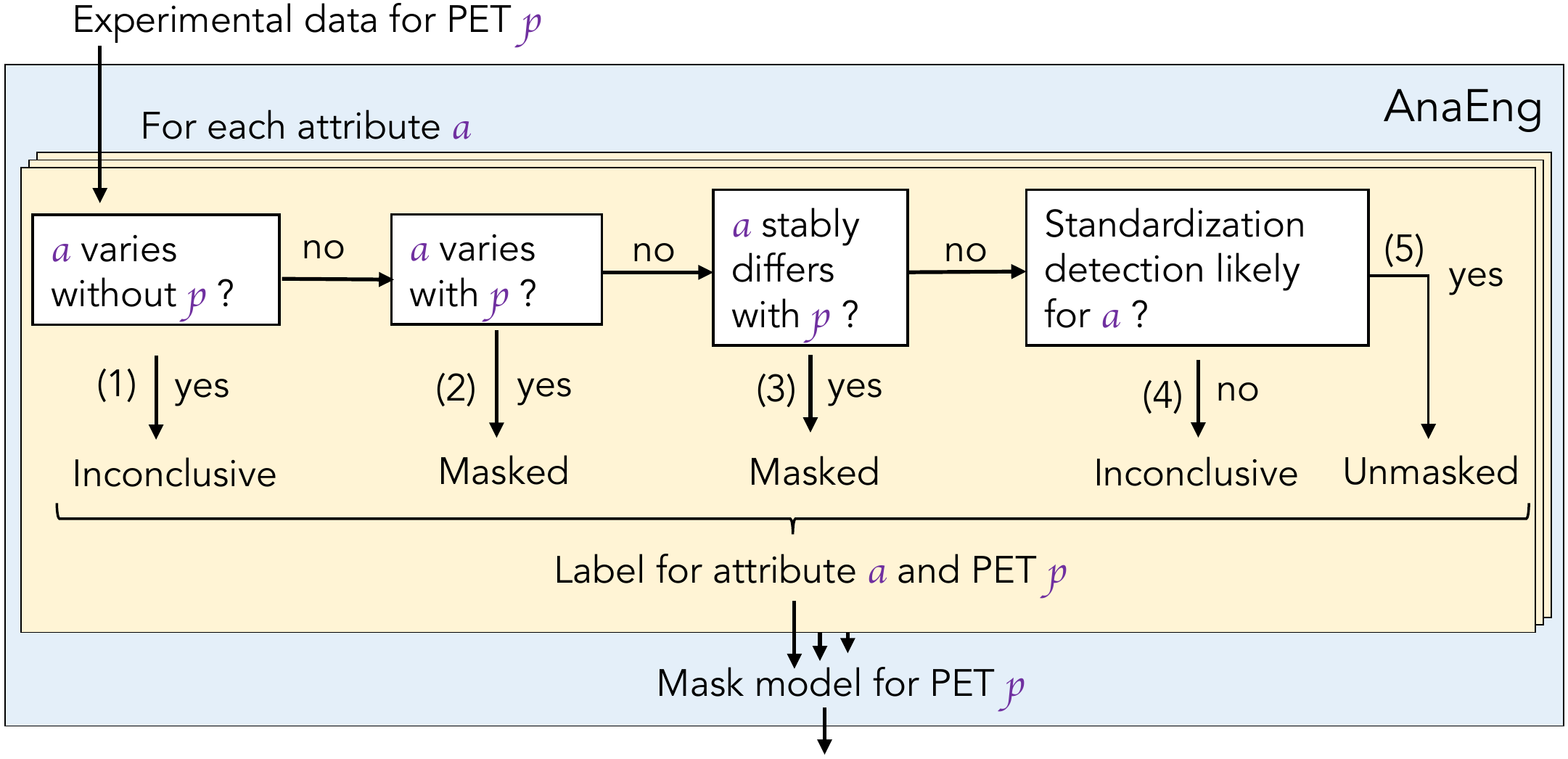}
\caption{The Analysis Engine (\ae) of \tool}
\label{fig:analysis}
\end{figure*}

\paragraph{Analysis Engine}

To check for masking by a PET, \ae{} uses both the fingerprints
collected by \fs{} from the browsers driven by \cs{} and information
directly from \cs{} stating which browser used which PETs and in which
configurations. 
Figure~\ref{fig:analysis} provides an overview of \ae{}. 
In short, the analysis looks for both masking by standardization and
by variation.
If it detects standardization or variation for an attribute, it models
the attribute as masked in the mask model of the PET that it produces.
It models an attributes as unmasked if it is able to thoroughly test
it and find neither type of masking.
The possible results of the analysis are
\begin{enumerate}
\item Label as inconclusive: variation testing impossible due the baseline browser varying the attribute
\item Label as masked: AFPET-induced variation detected
\item Label as masked: AFPET-induced standardization detected
\item Label as inconclusive: partial standardization cannot be ruled out due to not having browsing platforms that differ enough in the attribute
\item Label as unmasked: impactful standardization ruled out as unlikely
\end{enumerate}

In more detail, \ae{} consumes a list of experimental
results from \cs{} performed on a variety of browsing platforms.
For each tested pair of an attribute and a PET, \ae{} first checks whether its value varied 
for any of browser platforms without the PETs installed
as they cross the three boundaries mentioned above.  
If so, it cannot detect whether the PET varies that attribute since it
is already varying.
In this case, \ae{} labels the attribute as inconclusive and records the reason
for this conclusion. %
If not, \ae{} goes on to check whether the attribute's value varied
for any of the browsing platforms with the addition of the PET and, if
so, labels the attribute as masked for this reason.

If the attribute is not labeled under either variation
check, \ae{} checks whether the attribute was masked by standardization.
First, for each baseline browsing platform, it checks whether the value differs between
the baseline platform and the platform with the PET installed.
If so, \ae{} concludes that the PET standardized the attribute
since the only difference between the two platforms is the addition of
the PET and variation has already been ruled out.
If not, then we can rule out full standardization with certainty but not partial standardization.

In general, ruling out partial standardization with experiments
requires testing for all possible attribute values, a prohibitively expensive,
if not impossible, task for many attributes.
However, \ae{} can, in reasonable time and with reasonable confidence,
rule out \emph{impactful} partial standardization, that is,
standardization that affects at least a fraction $f$ of the values.
To do so, \ae{} estimates the probability of seeing at least one
changed value given that at least a fraction of them $f$ are
being standardized.
If this probability is below some threshold $\alpha$,
\ae{} rejects the idea that tool is impactfully standardizing
and labels the attribute as unmasked with confidence $\alpha$.
Otherwise, the result is inconclusive since
not enough values of the attribute were tested.
We use the geometric distribution to estimate likelihood
of finding masking given that a fraction $f$ is happening.
Appendix~\ref{app:stats} provides details.

\subsection{Experiment}

Using \tool{}, we performed an initial experiment finding no 
additional spoofing from AFPETs crossing sessions.
\spacecut{(See Appendix~\ref{app:session} for details.)}%
Thus, to save time, our main experiment uses only a single session and does
not check for the masking of attributes by variation across sessions. 

We use \cs\ to simulate six browsing platforms. 
Three of these are virtual machines running various versions of Linux. 
We introduce additional changes into these virtual machines to simulate differences 
in the system configurations. Specifically, we install different fonts and browser versions, 
set up different timezones, and simulate different screen resolutions and languages, 
The remaining platforms run natively on a Linux desktop, Macbook Pro, and a PC laptop.
We perform measurements on Firefox and Chrome browsers. 
More details on these configurations are in Table~\ref{tab:configs}
\begin{table*}[t]
\caption{Configurations of simulated browsing platforms in our main
experiment. 
The last three were regularly used.}
\label{tab:configs}
\centering\footnotesize
\setlength\tabcolsep{3pt}
\renewcommand{\arraystretch}{1.3} %
\begin{tabular}{@{}llllllllll@{}}
\toprule
\multirow{2}{*}{\#} & \multirow{2}{*}{Type} & \multirow{2}{*}{OS} &  Addl. & \multirow{2}{*}{Resolution} & Locale & \multirow{2}{*}{Timezone} & \multicolumn{2}{c}{Browser versions} & Notes\\
\cline{8-9}
&  &  &  Fonts &  &  \& LANG  &  & Firefox  & Chrome \\
\midrule
 $1$ &   VM  &  Ubuntu 16.04  &  Mordred &   $450{\times}721{\times}24$  &   ru\_RU.UTF-8  &   GMT+6  &   56.0  &   63.0 \\  
 $2$ &   VM  &  Debian 8.10  &  OldLondon  &   $2000{\times}2000{\times}16$  &   de\_DE.UTF-8  &   GMT-3  &   56.0  &   63.0 \\ 
 $3$ &   VM  &  Ubuntu 14.04  &  (none added)  &   $6000{\times}3000{\times}24{+}64$  &   ar\_SA.UTF-8  &   GMT-11  &   56.0  &   63.0 \\ 
 $4$ &   Local  &  Ubuntu 16.04 & $> 40$ & $1920{\times}1080{\times}24$ &   en\_EN.UTF-8  & GMT-8  & 56.0 & 70.0 \\
 $5$ &   Local  &  macOS 10.13  &  $> 145$  &   $1440{\times}900{\times}24$  &   en\_EN.UTF-8  &   GMT-8  &   56.0  &   70.0 \\ 
 $6$ &   Local  &  Windows NT 10.0& $> 145$ & $1280{\times}720{\times}24$ &   en\_EN.UTF-8  & GMT-8  & 56.0 beta & 69.0 & Touch screen\\
\bottomrule
\end{tabular}

\end{table*}

\cs\ drives these experimental browsing platforms to load \fs\ 
for five reloads of 
each of the two domain names of \fs{}.
For each platform, 
it does these reloads a total of $28$ times: one time each for $26$ PETs
and one time each for the two baseline browsers.
All PETs are left in their
default configuration.

\subsection{Results}\label{sec:effects}

Before commenting on PETs, we make some observations about the
baseline browsers.
While we did not think of the choice of browsers as affecting the trackability of fingerprints, 
it turns out that comparing our baseline measurements for the 
two browsers reveals small differences in the attributes shared by them. 
Among the simulated platforms, 
Chrome sets the $\mathtt{cpu\ class}$ to \emph{unknown}, 
the $\mathtt{screen.Depth}$ to $24$, and the $\mathtt{buildID}$ to \emph{Undefined}, 
unlike Firefox which reveals different values across browsing platforms.
On the other hand, Firefox does not reveal any plugins, while Chrome does. 
Chrome's plugins differ across Ubuntu, Debian, and macOS. 
\spacecut{Table~\ref{tab:basevals} in Appendix~\ref{app:pettables}
provides more details on the attribute values 
revealed by each baseline browser.}%
\tool{} does not find any baseline browser to vary any attributes itself
(outcome (1) in Fig.~\ref{fig:analysis}).

Turning to PETs, \tool{} automatically produces Table~\ref{tab:studybullet} which 
displays attributes masked or not by AFPETs. We comment upon the BLPETs in text. 
We provide \tool{} with the masks that each tool's documentation
purports, which it uses to facilitate comparing documented behaviors
with observed behaviors.
\spacecut{Table~\ref{tab:study} in Appendix~\ref{app:pettables} provides a lower level
description of the finding in terms of the type of masking found.}%

\begin{table*}
\centering
\caption{AFPET masks as purported and observed by \tool{}. 
$\square$ indicates AFPET's documentation purports that the attribute is masked. 
The remaining symbols represent the possible outputs of \tool{}:
$\masked$ indicates observed masking, 
$\times$ indicates no masking found even when it is likely to detect it, and
$\cdot$ indicates inconclusive results. %
For the results that we manually double checked, we include the outcome of that check as a superscript.
Here, $\color{blue}\times$ denotes that the PET really does not mask the attribute, $\color{blue}\masked$ that it does, $\color{blue}\times / \masked$ that it is not masked by default but can be with configuration,
and $\color{blue}?$ that the manual analysis was inconclusive.
Not shown are attributes that had all inconclusive results and not purported masking (nothing but $\cdot$):
$\mathtt{DNT\ enabled}$,
$\mathtt{IE\ addBehavior}$,
$\mathtt{adBlock\ installed}$,
$\mathtt{h.Connection}$,
$\mathtt{h.Dnt}$,
$\mathtt{h.Up.{\mhyphen}Ins.{\mhyphen}Req.}$,
$\mathtt{indexedDB}$,
$\mathtt{math.acosh(1e300)}$,
$\mathtt{math.asinh(1)}$,
$\mathtt{math.atanh(05)}$,
$\mathtt{math.cbrt(100)}$,
$\mathtt{math.cosh(10)}$,
$\mathtt{math.expm1(1)}$,
$\mathtt{math.log1p(10)}$,
$\mathtt{math.sinh(1)}$,
$\mathtt{math.tanh(1)}$,
and $\mathtt{openDB}$.}
\label{tab:studybullet}
\setlength\tabcolsep{5pt}
\renewcommand{\arraystretch}{1.22} %
\begin{tabular}{@{}lllllllllllllllll@{}}
\toprule
& \multicolumn{7}{c}{Chrome} && \multicolumn{8}{c}{Firefox} \\ 
\cline{2-8} \cline{10-17}
Attribute  &  \Sbr  &  \Scdc  &  \Scfb  &  \Sgl  &  \Shmf  &  \Spe  &  \Str  &  &  \Sbi  &  \Sbl  &  \Scdf  &  \Sca  &  \Snee  &  \Ssf  &  \Stor  &  \Sto  \\ 
\midrule
$\mathtt{buildID}$  & $\inconc$ & $\inconc$ & $\inconc$ & $\inconc$ & $\inconc$ & $\inconc$ & $\inconc$ &  & $\squarecheck$ & $\squarecheck$ & $\times$ & $\times$ & $\times$ & $\times$ & $\squarecheck$ & $\masked$ \\ 
$\mathtt{canvas\ fingerprint}$  & $\squarecheck$ & $\squarecheck$ & $\squarecheck$ & $\squarecheck$ & $\squarecheck$ & $\squaretimes^{\color{blue}\times/\masked}$ & $\squaretimes^{\color{blue}\times}$ &  & $\times$ & $\times$ & $\squarecheck$ & $\squarecheck$ & $\times$ & $\times$ & $\squarecheck$ & $\times$ \\ 
$\mathtt{cookies\ enabled}$  & $\inconc$ & $\inconc$ & $\inconc$ & $\inconc$ & $\inconc$ & $\squaredot$ & $\inconc$ &  & $\inconc$ & $\inconc$ & $\inconc$ & $\inconc$ & $\inconc$ & $\inconc$ & $\inconc$ & $\inconc$ \\ 
$\mathtt{cpu\ class}$  & $\inconc$ & $\inconc$ & $\inconc$ & $\inconc$ & $\inconc$ & $\inconc$ & $\inconc$ &  & $\squarecheck$ & $\squarecheck$ & $\times$ & $\times$ & $\times$ & $\masked$ & $\squarecheck$ & $\masked$ \\ 
$\mathtt{h.Accept}$  & $\inconc$ & $\inconc$ & $\inconc$ & $\inconc$ & $\inconc$ & $\squaredot$ & $\inconc$ &  & $\inconc$ & $\inconc$ & $\inconc$ & $\inconc$ & $\inconc$ & $\inconc$ & $\squaredot$ & $\inconc$ \\ 
$\mathtt{h.Accept{\mhyphen}Encoding}$  & $\inconc$ & $\inconc$ & $\inconc$ & $\inconc$ & $\inconc$ & $\inconc$ & $\inconc$ &  & $\inconc$ & $\inconc$ & $\inconc$ & $\inconc$ & $\inconc$ & $\inconc$ & $\squaredot$ & $\inconc$ \\ 
$\mathtt{h.Accept{\mhyphen}Language}$  & $\masked$ & $\times$ & $\times$ & $\times$ & $\times$ & $\times$ & $\times$ &  & $\times$ & $\squarecheck$ & $\times$ & $\times$ & $\times$ & $\times$ & $\squarecheck$ & $\times$ \\ 
$\mathtt{h.Pragma}$  & $\inconc$ & $\inconc$ & $\inconc$ & $\inconc$ & $\masked$ & $\inconc$ & $\inconc$ &  & $\inconc$ & $\inconc$ & $\inconc$ & $\inconc$ & $\inconc$ & $\inconc$ & $\inconc$ & $\inconc$ \\ 
$\mathtt{h.User{\mhyphen}Agent}$  & $\squarecheck$ & $\times$ & $\times$ & $\times$ & $\squarecheck$ & $\squaretimes^{\color{blue}\times/\masked}$ & $\squaretimes^{\color{blue}\times}$ &  & $\squarecheck$ & $\squarecheck$ & $\times$ & $\times$ & $\times$ & $\times$ & $\squarecheck$ & $\squarecheck$ \\ 
$\mathtt{javascript\ fonts}$  & $\times$ & $\times$ & $\times$ & $\times$ & $\times$ & $\times$ & $\times$ &  & $\times$ & $\times$ & $\times$ & $\times$ & $\times$ & $\squaretimes^{\color{blue}?}$ & $\squarecheck$ & $\times$ \\ 
$\mathtt{language}$  & $\masked$ & $\times$ & $\times$ & $\times$ & $\times$ & $\times$ & $\times$ &  & $\times$ & $\squarecheck$ & $\times$ & $\times$ & $\times$ & $\times$ & $\squarecheck$ & $\times$ \\ 
$\mathtt{local\ storage}$  & $\inconc$ & $\inconc$ & $\inconc$ & $\inconc$ & $\inconc$ & $\squaredot$ & $\inconc$ &  & $\inconc$ & $\inconc$ & $\inconc$ & $\inconc$ & $\inconc$ & $\inconc$ & $\inconc$ & $\inconc$ \\ 
$\mathtt{platform}$  & $\times$ & $\times$ & $\times$ & $\times$ & $\times$ & $\times$ & $\times$ &  & $\squarecheck$ & $\squarecheck$ & $\times$ & $\times$ & $\times$ & $\times$ & $\squaretimes^{\color{blue}\times}$ & $\masked$ \\ 
$\mathtt{plugins}$  & $\squarecheck$ & $\times$ & $\times$ & $\times$ & $\times$ & $\times$ & $\times$ &  & $\inconc$ & $\inconc$ & $\inconc$ & $\inconc$ & $\squaredot^{\color{blue}+}$ & $\squaredot^{\color{blue}+}$ & $\squaredot$ & $\inconc$ \\ 
$\mathtt{screen.AvailHeight}$  & $\times$ & $\times$ & $\times$ & $\times$ & $\times$ & $\times$ & $\times$ &  & $\times$ & $\times$ & $\times$ & $\times$ & $\times$ & $\squarecheck$ & $\squarecheck$ & $\times$ \\ 
$\mathtt{screen.AvailLeft}$  & $\times$ & $\times$ & $\times$ & $\times$ & $\times$ & $\times$ & $\times$ &  & $\times$ & $\times$ & $\times$ & $\times$ & $\times$ & $\squarecheck$ & $\squarecheck$ & $\times$ \\ 
$\mathtt{screen.AvailTop}$  & $\times$ & $\times$ & $\times$ & $\times$ & $\times$ & $\times$ & $\times$ &  & $\times$ & $\times$ & $\times$ & $\times$ & $\times$ & $\squarecheck$ & $\squarecheck$ & $\times$ \\ 
$\mathtt{screen.AvailWidth}$  & $\times$ & $\times$ & $\times$ & $\times$ & $\times$ & $\times$ & $\times$ &  & $\times$ & $\times$ & $\times$ & $\times$ & $\times$ & $\squarecheck$ & $\squarecheck$ & $\times$ \\ 
$\mathtt{screen.Depth}$  & $\inconc$ & $\inconc$ & $\inconc$ & $\inconc$ & $\inconc$ & $\inconc$ & $\inconc$ &  & $\times$ & $\times$ & $\times$ & $\times$ & $\times$ & $\squarecheck$ & $\squarecheck$ & $\times$ \\ 
$\mathtt{screen.Height}$  & $\times$ & $\times$ & $\times$ & $\times$ & $\times$ & $\times$ & $\times$ &  & $\times$ & $\times$ & $\times$ & $\times$ & $\times$ & $\squarecheck$ & $\squarecheck$ & $\times$ \\ 
$\mathtt{screen.Left}$  & $\inconc$ & $\inconc$ & $\inconc$ & $\inconc$ & $\inconc$ & $\inconc$ & $\inconc$ &  & $\inconc$ & $\inconc$ & $\inconc$ & $\inconc$ & $\inconc$ & $\squaredot$ & $\squaredot$ & $\inconc$ \\ 
$\mathtt{screen.Pixel\ Ratio}$  & $\times$ & $\times$ & $\times$ & $\times$ & $\times$ & $\times$ & $\times$ &  & $\times$ & $\times$ & $\times$ & $\times$ & $\times$ & $\squarecheck$ & $\squarecheck$ & $\times$ \\ 
$\mathtt{screen.Top}$  & $\inconc$ & $\inconc$ & $\inconc$ & $\inconc$ & $\inconc$ & $\inconc$ & $\inconc$ &  & $\inconc$ & $\inconc$ & $\inconc$ & $\inconc$ & $\inconc$ & $\squaredot$ & $\squaredot$ & $\inconc$ \\ 
$\mathtt{screen.Width}$  & $\times$ & $\times$ & $\times$ & $\times$ & $\times$ & $\times$ & $\times$ &  & $\times$ & $\times$ & $\times$ & $\times$ & $\times$ & $\squarecheck$ & $\squarecheck$ & $\times$ \\ 
$\mathtt{session\ storage}$  & $\inconc$ & $\inconc$ & $\inconc$ & $\inconc$ & $\inconc$ & $\squaredot$ & $\inconc$ &  & $\inconc$ & $\inconc$ & $\inconc$ & $\inconc$ & $\inconc$ & $\inconc$ & $\inconc$ & $\inconc$ \\ 
$\mathtt{timezone}$  & $\times$ & $\times$ & $\times$ & $\times$ & $\times$ & $\times$ & $\times$ &  & $\times$ & $\times$ & $\times$ & $\times$ & $\times$ & $\times$ & $\squarecheck$ & $\times$ \\ 
$\mathtt{touch.event}$  & $\inconc$ & $\inconc$ & $\inconc$ & $\inconc$ & $\inconc$ & $\inconc$ & $\inconc$ &  & $\times$ & $\times$ & $\times$ & $\times$ & $\times$ & $\times$ & $\squarecheck$ & $\times$ \\ 
$\mathtt{touch.max\ points}$  & $\times$ & $\times$ & $\times$ & $\times$ & $\times$ & $\times$ & $\times$ &  & $\inconc$ & $\inconc$ & $\inconc$ & $\inconc$ & $\inconc$ & $\inconc$ & $\inconc$ & $\inconc$ \\ 
$\mathtt{touch.start}$  & $\inconc$ & $\inconc$ & $\inconc$ & $\inconc$ & $\inconc$ & $\inconc$ & $\inconc$ &  & $\times$ & $\times$ & $\times$ & $\times$ & $\times$ & $\times$ & $\masked$ & $\times$ \\ 
$\mathtt{webGL.Data\ Hash}$  & $\squarecheck$ & $\masked$ & $\masked$ & $\masked$ & $\masked$ & $\times$ & $\times$ &  & $\times$ & $\times$ & $\masked$ & $\masked$ & $\times$ & $\times$ & $\squarecheck$ & $\times$ \\ 
$\mathtt{webGL.Renderer}$  & $\squarecheck$ & $\masked$ & $\masked$ & $\masked$ & $\masked$ & $\times$ & $\times$ &  & $\times$ & $\times$ & $\masked$ & $\times$ & $\times$ & $\times$ & $\squarecheck$ & $\times$ \\ 
$\mathtt{webGL.Vendor}$  & $\squarecheck$ & $\masked$ & $\masked$ & $\masked$ & $\masked$ & $\times$ & $\times$ &  & $\times$ & $\times$ & $\masked$ & $\times$ & $\times$ & $\times$ & $\squarecheck$ & $\times$ \\ 
\bottomrule
\end{tabular}

\end{table*}

Among the $15$ AFPETs, %
three (\tr, \pe, and \nee) do not lead to any detectable masking
in their default configurations. %
The remaining $12$ AFPETs mask at least one of the collected attributes.  

Our experiment also detects undocumented masking of attributes by AFPETs. 
For example, while \cdc, \cdf, \cfb, \gl, and \ca\ claim to spoof only the 
$\mathtt{canvas\ fingerprint}$, we also find them spoofing $\mathtt{webGL}$ attributes. 
Similarly, we find undocumented modifications by \br, \sf, and \to. 
We also find inconsistencies in the behavior of \br, \pbc, \pbf, \hmf, and \tor{}, which 
we discussed in Section~\ref{sec:intro-results}.

Among the $11$ BLPETs,
four (\df, \dc, \ghc, and \ghf) do not lead to any
detectable modifications of attributes, four (\apc, \apf, \uoc, and \uof) modify the attribute 
$\mathtt{adBlock\ installed}$, and three (\pbc, \pbf, and \tp) modify Do Not Track attributes. 
As discussed in the introduction, these BLPETs are presented as
AFPETs, but their modifications can actually make their users more
identifiable.
Indeed, Privacy Badger was updated in response to our finding.

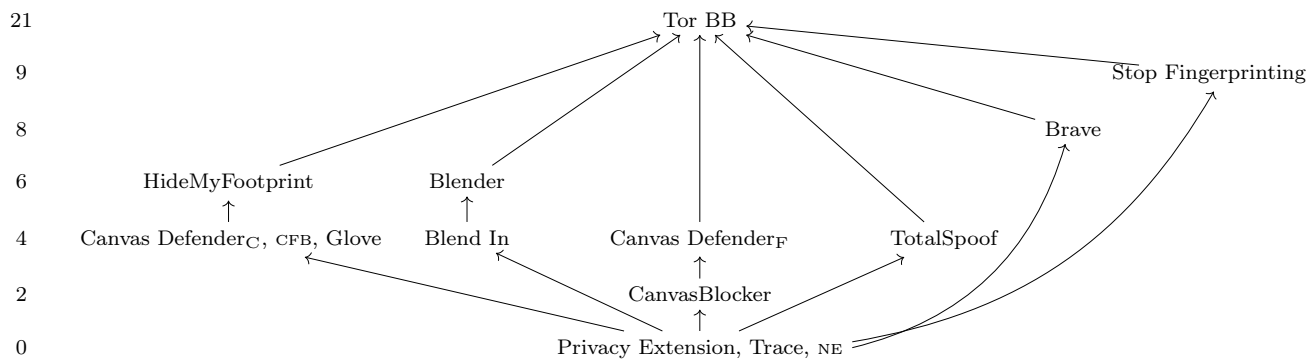
\begin{figure*}[t]
\begin{center}
\footnotesize
\begin{tikzcd}[math mode=false, row sep=1em, column sep=1.3em]
21 & & & \tor \\
9  & & & & & & \mbox{}\hspace{-4ex}\sf\arrow[ulll] \\
8  & & & & & \br\arrow[uull] \\
6  & \hmf\arrow[uuurr] &  \bl\arrow[uuur] \\ %
4  & \cdc, \Scfb, \gl\arrow[u]\hspace{-0.5ex} & \bi\arrow[u] & \cdf\arrow[uuuu] & \to\arrow[uuuul] \\ %
2  & & & \ca\arrow[u] \\ %
0  & & & \pe, \tr, \Snee\hspace{-30ex}\arrow[uull]\arrow[uul]\arrow[u]\arrow[uur]\arrow[uuuurr, bend right=30]\arrow[uuuuurrr, bend right=25] \\ %
\end{tikzcd}
\end{center}
\caption{Ranking of AFPETs by the experimental method.  
Arrows show the pre-order, with AFPETs at an equivalent order being grouped together. 
The y-axis shows the number of attributes masked.}
\label{fig:expr-ranking}
\end{figure*}

Given the difficulty of taking in Table~\ref{tab:studybullet}, 
for the purpose of ranking AFPETs relative to one another, we provide 
Figure~\ref{fig:expr-ranking} considering each of these masking behaviors 
as equally valuable for reducing trackability. 
This level of abstraction in modeling AFPETs seems reasonable given our
belief that trackers are foiled by any of these methods given the
complexity of, for example, using a varying attribute for tracking.
We produce a pre-ordering of the AFPETs where one AFPET $\P_1$ is above or
equal to another AFPET $\P_2$ iff $\P_1$ masks every attribute that
$\P_2$ does.
Those desiring a finer gradation can look at the number of attributes
masked, but must bear in mind that not all attributes are equally important to mask. 
The hybrid evaluation discussed in Section~\ref{sec:datadriven} takes into account 
the relative importance of different for ranking AFPETs. 

\subsection{Discussion and Limitations}\label{sec:explimitations}

As discussed in Section~\ref{sec:intro-interpretation}, the ranking
above may not be suitable for some evaluation goals.
For example, some AFPETs were designed to mask a single attribute and
does in fact mask it (e.g., \cdc).
Our findings that such AFPETs (or BLPETs) do not mask all
attributes should not be interpreted as the PET having a bug.
Nevertheless, consumers and advocates seeking effective
PETs may find our results useful.

As mentioned above, we may miss some masking of attributes due to not testing values that a PET standardizes away.
Furthermore, we may not detect a PET varying an attribute across a boundary that we do not test.
Thus, while we can be sure of masking when we find it, we cannot be sure we have found all masking.

\fs\ extracts fingerprints by running 
first-party fingerprinting scripts on browsing platforms. 
Thus, we do not detect masking that happens for only
third-party scripts.

To an extent, these limitations can be mitigated with more 
comprehensive experiments using \tool. For example, one can modify \fs\ to collect 
additional attributes in both first-party and third-party contexts.
Moreover, one can modify \cs\ to detect variations across other boundaries
and use more diverse experimental browsing platforms to be more confident about not missing 
standardization modifications. We will make \tool\ freely available for 
more extensive experimentation and further development. 
Our current evaluations demonstrate the benefits
of an experimental evaluation method for AFPETs within the current boundaries. 

Where our experiments dispute claimed masking ($\squaretimes$ in
Table~\ref{tab:studybullet}), it may be due to the above limitations 
rather than documentation making spurious claims. 
PETs may mask
more attributes when configured to do so, but users find
it difficult to change the defaults~\cite{leon2012johnny},
suggesting our experiments may capture typical use.
Next, we perform additional manual analysis 
to understand
the effects of configuration and why our results are in conflict with the
documentation of some PETs.

\subsection{Additional Manual Analysis}
\label{sec:manual}

To address some of the limitations mentioned above, we perform
manual analysis of some PETs. Specifically, we analyze AFPETs 
for which we found no evidence of any masking (\tr{}, \pe{}, \nee{}) and those which 
made a claim rejected by \tool{} (\tr{}, \pe{}, \sf{}, Tor).

\tool{} rejects two claims of 
masking by \tr{}. We could not find the source code for \tr, 
but we installed the extension and manually examined it.
Both the documentation and settings panel show canvas fingerprinting
being masked by default, despite our studies concluding the opposite.
As far as we can tell, \tr{} really does not mask this attribute
despite claiming to. Since running our tests, \tr{} has been updated from
version 1.0.2 to 1.8.6 and it now randomizes the canvas fingerprinting.

As for masking the user-agent, the settings panel of \tr{} shows that user-agent randomization is off by
default, explaining our finding.
(It's under the ``Advanced Features'' tab, despite their 
	webpage prominently advertising the feature.)
Turning it on does randomize the user agent.

All of \pe{}'s masking abilities are off by default.
Turning them on does result in standardizing the two attributes in
question: the canvas fingerprint and the user-agent.

To analyze \nee{}, we examined both its source code and documentation.
The documentation of \nee{} only claims to mask plugins and we found
evidence of plugin masking in \nee{}'s source code.
\tool{} was inconclusive for this attribute since it was unable to
exercise the plugin list for Firefox due to Firefox making the loading
of any plugins a manual process. %
Thus, this instance does not represent a false negative, and instead represents a
failure to find a positive.

We manually tested \sf{} and found that, like \nee{}, it masks
plugins despite \tool's inconclusive finding.
As for the rejected claim of masking javascript fonts, \sf{} may
be doing something with the fonts, but not enough to defeat
the way \fs{} fingerprints them.

Examining \tor{} leads us to believe that a recent update (after Version~7.0.11), accidentally affected its masking of the $\mathtt{platform}$
attribute.
We also find that during the same time frame, the $\mathtt{cpu class}$
and $\mathtt{h.User{\mhyphen}Agent}$ went from being apparently fully masked
to partially masked.
We have found user complaints about this change in
behavior for version~8.0a10~\cite{anonymous18torblogcomment}.

We also confirmed that Privacy Badger did not set the \textsf{doNotTrack} field
of the navigator object to match the \textsf{Dnt} header.  
The code was fixed after we
notified the developers of the issue~\cite{ghostwords18pull}.

\section{Hybrid Evaluation of AFPETs}
\label{sec:datadriven}

Our experimental method provides a model of how various AFPETs mask fingerprints
as well a ranking of AFPETs based on the number of attributes they mask. 
However, it does not consider how important masking each attribute
is.

We develop a hybrid method that combines the benefits of the experimental
method with an observational method. 
We start by considering a completely observational method and then
discuss how combining it with our experimental method allows us to
overcome each of their limitations.

In short, the method uses
mask model of each AFPET provided by the experimental method.
For each attribute, we model the AFPET as masking the attribute 
if the mask model indicates so or if the experiment was inconclusive, 
thereby overestimating the AFPET's abilities. We use this mask model to transform 
a set of original fingerprints collected on the \texttt{amiunique} website 
into a counterfactual AFPET-modified set, which simulates the browsing platforms 
in the original dataset visiting \texttt{amiunique} with an AFPET installed. 
To determine the effectiveness of the AFPET, we compare trackability in the two datasets.

\subsection{Sampling}

We cannot, in practice, see all the world's browsing platforms 
and instead must work with a sample. The quality of the metrics computed from the sample 
depends upon both the nature of the metric and the sample.
For example, a random sample will provide a reasonable estimation of the entropy 
(e.g.,~\cite{paninski2003estimation}). %
However, estimating the proportion of users in small anonymity sets from even a 
random sample proves difficult since the length of the
tail of the distribution may be unclear from a random sample.

Furthermore, in practice, we must approximate truly random samples of
browser platforms from available datasets since we cannot force all users to participate.
We do so by using a convenience sample provided to us by the
\texttt{amiunique} website, which collects fingerprints to understand how trackable they are. 
This sample comprises 25,984 
real-world fingerprints collected over a period of $30$ days (10/02/2017 to 11/02/2017). 
Each fingerprint comprises $32$ different attributes.
\spacecut{A full list is in Table~\ref{tab:aiuattributes}
in Appendix~\ref{app:pettables}.}%

Determining the representativeness of this sample 
is difficult since it can only be compared to other possibly unrepresentative samples.
We compare our sample's distributions to GlobalStat's for desktop
users~\cite{statcounter}.
We find that our sample has a higher proportion of Firefox users
(42\% vs.\ 12\%) and of Linux users (19\% vs.\ 1\%).
Perhaps people visiting browser fingerprinting websites have more
technical knowledge and a preference for open-source technologies.
\spacecut{Table~\ref{tbl:compare-samples} in Appendix~\ref{app:pettables} provides details.}%

\subsection{Metrics of Trackability}
\label{sec:metrics}

We haven't yet defined what we mean 
by \emph{trackability}. Is a tracker that can determine with 10\% certainty 90\% of the time 
that you visited a website worse than one that can determine it with 90\% 
certainty 10\% of the time?
This depends upon both the tracker's and the evaluator's goals.
With this in mind, we do not argue for a single metric, but rather consider a few.

To measure trackability of the fingerprints, 
we have implemented \otool, which consumes a dataset
and characterizes how trackable its members are. One such characterization is the 
anonymity set. An \emph{anonymity set} comprises browsing platforms with
identical fingerprints that are, thus, indistinguishable from each other. 
Thus, the smaller and more numerous the anonymity sets, the higher the uniqueness.
\otool{} implements 
various proposed functions of the distribution of anonymity 
sets of browsing platforms for measuring 
uniqueness~\cite{yen2012host, eckersley2010unique}.

The first metric which we use to measure uniqueness is entropy. 
For a set of browser platforms $\D = \{b_i\}_i$, such as those using a
particular AFPET, let $\D[id(\cdot)]$ denote the multiset of fingerprints
$\{id(b_i)\}_i$ where $id(\cdot)$ is the fingerprinting mechanism.
The entropy of these fingerprints is given by 
\[ \entropy(\D[id(\cdot)]) = {-}\sum_{id_k \in \D[id(\cdot)]}\Pr[id_k]\log_2(\Pr[id_k]) \]
where $\Pr[id_k]$ is the probability of 
observing the fingerprint $id_k$, which we estimate from the frequency of $id_k$
in $\D[id(\cdot)]$. The higher the entropy, the higher the uniqueness of the fingerprints.

\otool{} also measures the proportion of users in anonymity sets of size
less than or equal to $1$ ($\pctal$) and $10$ ($\pctbl$). 
These metrics measure the proportion of 
browsing platforms hiding in anonymity sets of sizes at most $1$ and $10$. 
The higher $\pctal$ is, the higher is the fraction of browsing platforms that can be 
uniquely identified. Similarly, higher $\pctbl$ indicates a higher fraction of 
browsing platforms that can be identified to a set of size at most $10$. 
Thus, higher values of these metrics indicate higher uniqueness of the fingerprints. 

\otool{} measures effectiveness of a PET $\P$ against fingerprinting mechanism 
$id(\cdot)$ 
from the dataset of fingerprints $\D[id(\cdot)]$
in terms of a metric $\func$ in $\{\entropy, \pctal, \pctbl\}$ 
as
\begin{align}
\eff_{\func}(\P, id, \D_{\P}, \D_{\bar{\P}}) &:= 
\func(\D_{\bar{\P}}[id(\cdot)])-\func(\D_{\P}[id(\cdot)])
\label{eqn:eff}
\end{align}
where $\D_\P$ is a subset of $\D$ using the PET and $\D_{\bar{\P}}$ is the rest of $\D$.

\subsection{Limitations of Observations Alone}

In principle, a highly-context dependent, completely observational method could function by
comparing the fingerprints produced by users of 
each AFPET to determine which are the least trackable.
In practice, 
we face difficulties with obtaining a representative sample of AFPET users
and determining which users run which AFPETs.

\textit{PET determination.}
Determining PET use from fingerprints not explicitly containing the information is 
difficult. This limitation can be overcome by a fingerprinting server designed
to collect information about PET use.
One approach is to ask visitors about their 
PETs, but users can be unaware of their own browser's configurations. 
In some cases, PETs have a distinctive fingerprint that gives away their use, 
but this would only help us with a subset of PETs. Moreover, this approach would not work 
with AFPETs which attempt to have a common fingerprint also had by non-users of the AFPET.
Alternatively, fingerprint collection websites can use automated methods to detect 
browser extension PETs (e.g.,~\cite{starov2017xhound,rola2017extension}).
Unfortunately, our observational data lacks this information.

\textit{PET sampling.}
Even with a fingerprinting server collecting PET information, 
getting a representative sample of real users 
with AFPETs to visit the website may be difficult, since there are few AFPET users. 
This is especially true for new and not yet popular AFPETs. Furthermore, users of AFPETs may 
be systematically different from users without AFPETs, thereby introducing confounding factors
influencing the trackability metrics. To remove or minimize the effect of these confounding
factors, one may have to identify matched pairs of users, one using an AFPET and another not.

Due to these limitations, we cannot apply \otool{} directly to our dataset. 
Moreover, the PET sampling limitation may prevent application of this method directly to
data collected on even fingerprinting servers designed for PET determination.
Thus, we instead use \otool{}
in an hybrid evaluation method that avoids the PET determination and sampling
problems altogether.

\subsection{Overcoming the Limitations of Observations}

To overcome the difficulty of getting a sample $\D_{\P}$ of browser platforms using a PET $\P$,
we construct our own from a sample $\D_{\bar{\P}}$ of browser platforms not using $\P$.
We then provide both to \otool{}, which uses
$\eff_{\func}$ (see Eq.~\ref{eqn:eff} above) to evaluate the PET $\P$,
as show in Figure~\ref{fig:method}.

This approach requires that we first get a sample of platforms not using $\P$.
We start with
the \texttt{amiunique} dataset.
To convert that dataset of fingerprints into one of platforms, we need a mapping of fingerprints to unique browsing platforms. 
We approximate this mapping using cookie IDs associated with each fingerprint. 
We treat fingerprints with different 
cookie IDs as being produced by different browsing platform. 
This approach is similar to Eckersley, who also uses cookies in 
his Panopticlick study to approximate returning visitors~\cite{eckersley2010unique}. 
In the dataset, 21,395 fingerprints have a cookie associated with them, of which, 18,295
are unique. 

To obtain $\D_{\bar{\P}}[id(\cdot)]$, we sanitize the dataset to remove fingerprints 
with obvious signs of PET use, specifically those with JavaScript disabled and 
illegitimate screen resolutions. 
Additionally, we only retain fingerprints from desktop browsers (with Windows, Mac, or Linux
OSes) since all the PETs we study are for desktops. These sanitizations 
leave 17,109 fingerprints. Finally, we separate this set into two sets of fingerprints,
one from Chrome and another from Firefox browsers by looking at the
\ua\ attribute in each fingerprint. 
This results in 9,493 Chrome and 6,516 Firefox browser
fingerprints, which we use to simulate the tracker's 
view of the original fingerprints for evaluating Chrome and Firefox AFPETs respectively. 
We find that the original fingerprints reveal $13.002$ and $12.359$ 
bits of entropy for Chrome and Firefox browsers respectively. These and other metrics 
are presented in Table~\ref{tab:summaries} corresponding to the `no mask' row. 

The mask model from the experimental method provides a way to transform these original 
fingerprints. 
We apply the mask model $\hat{\P}$ of an PET $\P$ produced by \tool{} to 
the sample $\D_{\bar{\P}}$ of platforms without a PET to generate 
a sample of fingerprints $\D_{\hat{\P}}[id(\cdot)]$.
This generated sample estimates what the original fingerprints
would had looked like had the platforms used the PET $\P$.
We  use \otool{} to calculate the trackability metrics of the modified fingerprints and unmodified fingerprints. 
By comparing the metrics 
of the original and $\hat{\P}$-modified fingerprints, we estimate the effectiveness of the PET $\P$.

Of the $49$ original attributes, \tool\ provides a conclusive characterization for 
$18$ attributes on Chrome browsers and $20$ attributes on Firefox browsers. 
Of these, only $12$ appear in the \texttt{amiunique.org} dataset.
For a given PET, we mask these $12$ attributes according to the model generated 
by \tool\ and fully mask the remaining $16$ attributes in the \texttt{amiunique.org} dataset 
for which the experiment is inconclusive. 
By fully masking inconclusive attributes, we overestimate the effectiveness of PETs.
Thus, we generate a set of PET-modified fingerprints (i.e., $\D_{\hat{\P}}[id(\cdot)]$) 
from the original fingerprints and measure effectiveness of the $15$ AFPETs. 

\subsection{Results}

We present the metrics of trackability from Section~\ref{sec:metrics} for both Chrome
and Firefox AFPETs in Table~\ref{tab:summaries}. 
The original fingerprints reveal $13.002$ and $12.359$ 
bits of entropy for Chrome and Firefox browsers respectively. Applying a
\emph{base mask}
comprising all inconclusive attributes reduces the entropies to $12.914$ and 
$12.177$ bits. 

\begin{table*}
\centering
\caption{Trackability metrics for AFPETs. 
}\label{tab:summaries}
\setlength\tabcolsep{6pt}
\renewcommand{\arraystretch}{1.2} %
\begin{tabular}{@{}lrrr@{}}
\toprule
PET &  $\entropy$  &  $\pctal$  &  $\pctbl$   \\ %
\midrule
\multicolumn{4}{c}{Chrome PETs}  \\
\midrule
no mask &  $13.002$  &  $0.892$  &  $0.983$  \\ 
base mask, \pe, \tr &  $12.914$  &  $0.829$  &  $0.982$  \\ 
\cdc, \Scfb, \gl &  $12.306$  &  $0.641$  &  $0.893$  \\ 
\hmf &  $11.77$  &  $0.497$  &  $0.825$  \\ 
\br  &  $8.108$  &  $0.072$  &  $0.262$  \\ 
\midrule
\multicolumn{4}{c}{Firefox PETs}  \\
\midrule
no mask &  $12.359$  &  $0.875$  &  $0.96$  \\ 
base mask, \nee &  $12.177$  &  $0.797$  &  $0.949$  \\ 
\bi, \to &  $12.049$  &  $0.747$  &  $0.936$  \\ 
\ca &  $12.002$  &  $0.7$  &  $0.941$  \\ 
\bl &  $11.875$  &  $0.678$  &  $0.924$  \\ 
\sf &  $11.778$  &  $0.726$  &  $0.919$  \\ 
\cdf &  $11.263$  &  $0.483$  &  $0.833$  \\ 
\tor &  $4.766$  &  $0.01$  &  $0.038$  \\ 
\bottomrule
\end{tabular}

\end{table*}

Our evaluations reveal that all AFPETs but \br\ and \tor\ reveal over $11$ bits of entropy 
and hence are marginally better than not using any AFPET at all. 
For these AFPETs, fewer than $20\%$ of the fingerprints  
are in anonymity sets of size greater than $10$. 
\br\ does better, leaking just over $8$ bits of entropy and 
having over $70\%$ of fingerprints in anonymity sets of size greater than $10$. 
\tor\ performs best since it modifies all the $12$ attributes we consider.

\subsection{Remaining Limitations}
\label{sec:datalimitations}

While this hybrid method enables us to perform a fine-grained evaluation of AFPETs with few users,
it inherits some of the limitations of the methods on which it builds.
For example, from a purely observational methods comes the limitations that 
samples can be biased and that no one metric can completely 
capture the quality of an AFPET.
From the experimental method of Section~\ref{sec:experiment}, we inherit the approximate nature of the mask model, 
which does not account for how attributes are masked and how that affects privacy. 

In particular, our analysis overestimates the effectiveness 
of all AFPETs, since we assume any modifications
of an attribute by an AFPET renders that attribute useless to a tracker. 
This may not be the case.
For example, \br\ spoofs the \ua\ and the \al\ headers to different values than Chrome. 
While these spoofed values may continue to reveal bits of entropy, we consider the attributes
to be rendered useless for tracking. Similarly, \tor\ also reveals
spoofed values of screen resolution. 

We can carry out a tighter evaluation by considering 
a tracker which can take advantage of the spoofed values. This evaluation requires
knowledge of how an AFPET spoofs the attribute. 
For \tor, we performed a manual code analysis to determine how
exactly \tor{} deals with screen resolution attributes.
We rerun the hybrid analysis on a hand crafted mask
model capturing this behavior instead of using the rough model produced by \tool{}.\footnote{
We create the handcrafted mask model of \tor{} from the  
Firefox patch at 
\url{https://gitweb.torproject.org/tor-browser.git/commit/?h=tor-browser-45.8.0esr-6.5-2&id=7b3e68bd7172d4f3feac11e74c65b06729a502b2}.}   
This provides a tighter evaluation for \tor{} that will serve as the basis for
our analysis in Section~\ref{sec:tradeoff}.

Finally, the above evaluations are performed on the same set of fingerprints and
applies the mask to every fingerprint in the dataset, simulating total adoption of the AFPET.
This approach is appropriate evaluations with a long-term prospective,
such as selecting an AFPET to fund,
since a properly promoted AFPET could become nearly universal in the future.
However, those looking to select a AFPET for usage today should be
concerned with the number of users each AFPET has since it will affect
the size of the anonymity set the AFPET produces.
In the next section, we consider a modification of the above method
for dealing with this issue.

\section{Adjusting for the Number of Users}
\label{sec:pop}

To observe the consequences of having user bases of different sizes, we also evaluate the  
AFPETs taking into account their popularity.
Ideally, we would do this by having the fingerprints all the users of a AFPET.
However, not having access to this set of fingerprints, we instead simulate them
by drawing random samples of fingerprints
from the \texttt{amiunique.org} dataset of size equal
to the number of AFPET users and estimate uniqueness metrics on the samples. 

Table~\ref{tab:popularity} %
displays the number of users of 
each AFPET in our list as of Dec.~2017. 
The popularity of extensions were obtained from the Firefox add-on library~\cite{mozilla17add-ons}
and the Chrome extensions webstore~\cite{google17chorem-web-store}.
Tor's popularity was obtained from the Tor Metrics webpage~\cite{tor17users}.
For AFPETs with an undisclosed number of users, such as \br and \tp,
we are unable to perform this evaluation.

\begin{table}
\centering
\caption{Popularity of PETs in our list as of Dec.~$2017$}
\renewcommand{\arraystretch}{1.3} %
\label{tab:popularity}
\begin{tabular}{@{}lr@{}}
\toprule
PET & \# users \\
\midrule
\multicolumn{2}{c}{Chrome} \\
\midrule
\apc & 10,000,000+\\
\br & NA \\
\cdc & 19,769\\
\cfb & 7,630\\
\gl & 342 \\
\hmf & 177 \\
\ghc & 2,788,951 \\
\pe & 915\\
\uoc & 10,000,000+ \\
\midrule
\multicolumn{2}{c}{Firefox} \\
\midrule
\apf & 13,760,128  \\
\bi & 858 \\
\bl & 1,816 \\
\cdf & 5,274 \\
\ca & 27,170  \\
\ghf & 1,064,473 \\
\sf & 1,754 \\
\tor & $\approx$4,000,000\\
\to & 265 \\
\tp & NA \\
\uof & 4,837,884 \\
\bottomrule
\end{tabular}
\end{table}

We also do not perform these evaluations for AFPETs with a user base greater 
than 17,109 (like \tor, \cdc{} and \ca), since we cannot draw a sample from our
dataset of sufficient size.
Attempting to draw such a sample by allowing the same fingerprint to
be sampled multiple times will overestimate the effectiveness of the
PET since such repeats will surely be in the same anonymity set even
for PETs that do nothing.

For all other AFPETs, we compute the mean and the standard error of mean 
($\mathsf{mean\pm sem}$) of the trackability metrics from $100$ random samples. 
Table~\ref{tab:popfull} displays the entropy-based effectiveness metrics for these AFPETs, 
sorted according to the effectiveness. We can see that \cfb\
scores better than \hmf\ due to its high popularity, contrary to 
the original evaluations in Table~\ref{tab:summaries}. 
We also see that the effectiveness of tools with identical effects increases 
with popularity. For example, \to{} and \bi{} both identically modify 
$12$ attributes, but \bi{} is more effective than \to{} due to its popularity. 
Table~\ref{tab:popfull} %
also provides estimates of the other 
trackability metrics for these AFPETs.

\begin{table*}
\centering
\caption{Uniqueness metrics for different AFPETs on samples scaled according to their 
popularity}\label{tab:popfull}
\renewcommand{\arraystretch}{1.2} %
\begin{tabular}{@{}lrrrr@{}}
\toprule
PET &  $\popularity$ &  $\entropy$ &  $\pctal$ &  $\pctbl$ \\ %
\midrule
\multicolumn{5}{c}{Chrome PETs}  \\
\midrule
\hmf &  $177.0$  &  $7.343$  &  $0.901$  &  $1.000$  \\ 
\gl &  $342.0$  &  $8.277$  &  $0.886$  &  $1.000$  \\ 
\cfb &  $7630.0$  &  $11.559$  &  $0.313$  &  $0.899$  \\ 
\midrule
\multicolumn{5}{c}{Firefox PETs}  \\
\midrule
\to &  $265.0$  &  $7.904$  &  $0.889$  &  $1.000$  \\ 
\bi &  $858.0$  &  $9.401$  &  $0.777$  &  $0.983$  \\ 
\sf &  $1754.0$  &  $9.994$  &  $0.641$  &  $0.939$  \\ 
\bl &  $1816.0$  &  $10.200$  &  $0.614$  &  $0.960$  \\ 
\cdf &  $5274.0$  &  $10.656$  &  $0.252$  &  $0.845$  \\ 
\bottomrule
\end{tabular}

\end{table*}

\section{Application:\\ Informing AFPET Design}
\label{sec:tradeoff}

With the ability to accept handcrafted mask models, our hybrid method 
can help AFPET developers make an informed choice while designing AFPETs.
By measuring the effectiveness of hypothetical designs, AFPET developers can compare different
masking strategies to tradeoff utility with trackability. We carry out such 
an exploration comparing alternate designs of \tor\ by applying our hybrid method on 
hypothetical \tor\ versions that mask attributes differently.
\tor\ leaks
some information about the screen resolution by only partially standardizing it.
Specifically, it resizes new browser windows in quanta (step/bucket sizes) of $200{\times}100$ pixels, 
while capping the window size at $1000{\times}1000$ pixels, and uses the client 
content window size as screen dimensions~\cite{torprivacy}. As a result all \tor\ users get 
placed into one of $50$ anonymity sets based on the revealed screen dimensions, 
as long as they do not change the window dimensions manually. 
We explore the impact of the choices of cap and quanta parameters 
on the effectiveness of \tor. 
We use the number of unutilized screen pixels due to 
a spoofing strategy as a measure of utility loss. We measure two variants: 
the total number of unutilized pixels (average absolute loss), as well as 
the number of unutilized pixels as a percentage of the 
available pixels (average percentage loss). Increasing the cap parameters 
and decreasing the quanta parameters reduces this loss. 
We first measure the effectiveness of alternate strategies with strictly lower utility loss (i.e., 
higher cap and lower quanta parameters) than \tor's. 
An exhaustive search of all 19,999 quanta parameters less than
\tor's (i.e., $200{\times}100$), while fixing the cap parameters 
at \tor's (i.e., $1000{\times}1000$), finds no strategy achieving higher effectiveness 
in all metrics than \tor. Similarly, fixing the quanta parameters at $200{\times}100$, while
increasing the cap parameters in steps of $50$ pixels from $1000{\times}1000$ to 
$2000{\times}2000$ does not uncover any strategy with higher effectiveness either. 
We perform these explorations on the Firefox fingerprints in the 
\texttt{amiunique.org} dataset.

Next, we explore strategies that trade-off losses resulting from one set of parameters (e.g., 
quanta) with gains from another (e.g., cap) with the goal of finding a strategy that 
reduces the utility loss while increasing the effectiveness. We find that cap width 
parameter to be the most in need of improvement since less than $13\%$ of \texttt{amiunique.org}
fingerprints have a screen width less than $1000$ pixels.
We consider alternative cap widths of $1350$, $1550$, and $1600$
since a higher percentage of fingerprints ($25\%$, $47\%$, and $51\%$ respectively) 
have screen widths less than these caps. 
We retain the cap height of $1000$ pixels as more than $50\%$ of the fingerprints 
remain below that cap. 
We exhaustively search for all 10,201 quanta in the range 
$200{\times}100$ to $300{\times}200$ for all three cap parameters. 
We set an upper bound of  $300{\times}200$ as the loss may be too high for
low-resolution displays for very high quanta parameters. We find $786$ and $291$ 
quanta parameters for cap widths of $1350$ and $1550$ respectively for which the 
losses are lower than \tor's, but the effectiveness is higher. 
We display strategies with the least quanta parameters in Table~\ref{tab:tradeoff}.
As we increase the cap width to $1600$, none of the quanta parameters
lead to a higher measure of effectiveness than \tor. 

\begin{table*}
\centering
\caption{Comparison of effectiveness and loss of \tor's original spoofing strategies 
with alternate strategies. }
\label{tab:tradeoff}
\setlength\tabcolsep{2pt}
\renewcommand{\arraystretch}{1.2} %
\begin{tabular}{@{}lllrrrrr@{}}
\toprule
Cap & Quanta && $\entropy$ & $\pctal$ & $\pctbl$ & Abs.\@ Loss & $\%$ Loss \\
\midrule
$1000{\times}1000$ & $200{\times}100$ && $2.902$ & $0.001$ & $0.010$ & $870k$ & $50.3\%$ \\
\midrule
 $1350{\times}1000$ & $200{\times}193$ && $2.715$ & $0.001$ & $0.009$ & $729k$ & $42.6\%$ \\
$1350{\times}1000$ & $269{\times}160$ && $2.901$ & $0.000$ & $0.009$ & $728k$ & $42.3\%$ \\
 $1550{\times}1000$ & $222{\times}197$ && $2.899$ & $0.000$ & $0.009$  & $666k$ & $40.3\%$ \\
 $1550{\times}1000$ & $295{\times}160$ && $2.882$ & $0.000$ & $0.010$ & $636k$ & $37.2\%$ \\
\bottomrule
\end{tabular}
\end{table*}

\section{Conclusion and Discussion}

We carry out an evaluation of $15$ different AFPETs against
fingerprinting using two different methods. 
We develop \tool\ and use it for experiments to determine how these PETs 
spoof 18--20 different attributes. 
In addition to uncovering inconsistencies, it provides 
a model of AFPETs' behaviors. While the experimental method provides
an evaluation in terms of the number of attributes that an AFPET masks, it cannot distinguish 
between the relative importance of masking different attributes. Our hybrid method leverages
a real-world fingerprinting dataset to provide a finer grained view into the impact
of modifying different attributes. 
We find \tor\ to be the most effective AFPET among the ones we evaluate using both methods. 
It standardizes the most attributes and reduces the trackability of revealed fingerprints 
by the highest margins among the AFPETs we evaluate. 
We also apply our hybrid method to find some hypothetical spoofing strategies which have a 
smaller utility loss than Tor, yet are just as effective. 

The Tor Project is part of the team behind the FPCentral fingerprinting
repository, which spans a comprehensive collection of fingerprinting techniques. 
This awareness helps \tor\ developers build comprehensive defenses against fingerprinting. 
This however does not mean that \tor\ users are protected against 
all possible fingerprinting attacks. 
Developers must be on the lookout for new fingerprinting techniques and build 
in fresh defenses.
While our tools cannot automatically invent new attributes, it can be
extended to test them, allowing an assessment for how to deal with
them.

We end with some suggestions for AFPET developers and evaluators.
We recommend that developers address any attribute that \tool{}
flags as unmasked.  The entropy results from our
hybrid method can aid in determining the order in which to address
various unmasked attributes. 
Given our experimental results, we expect this task 
will keep the developers of most AFPETs busy.
Next, they might want to consider any attributes that
\tool{} labeled as inconclusive. %
After addressing these attributes, 
they can consider improving how an AFPET spoofs an attribute.
As shown in Section~\ref{sec:tradeoff}, not all spoofing is equal.
Developers should consider using \tor{} as a starting point for their development and
carefully consider the default settings of their AFPET.

The
set of fingerprintable attributes are open ended and will never be
fully enumerated, but new attributes can be added to our tools.
AFPET evaluators should keep in mind that any one-time evaluation of PETs will quickly become out of date.
We encourage developers and advocates (e.g., the EFF) to use automated tools to
regularly test the trackability of PETs.  Our tool can fill this need.

\section*{Acknowledgements}

We thank Milan Ganai for investigating how to automate the use of PETs on Windows.
We thank Lay Kuan Loh and Zheng Zong for assistance in exploring the application 
of information flow experiments to evaluate PETs. 
We thank Anupam Datta for discussions about this work.
We gratefully acknowledge funding support from
the National Science Foundation
(Grants 1514509 %
and 1704985). %
The opinions in this paper are those of the authors and do not
necessarily reflect the opinions of any funding sponsor or the United
States Government.

\bibliographystyle{plainurlmct}
\bibliography{biblio}

\begin{thebibliography}{10}

\bibitem{hidemyfootprint}
{Absolute Double}.
\newblock {HideMyFootprint}: Protect your privacy.
\newblock \url{https://hmfp.absolutedouble.co.uk}, 2017.
\newblock Accessed Dec.~25, 2017.

\bibitem{trace}
{Absolute Double}.
\newblock {Trace}: Browse online without leaving a trace.
\newblock \url{https://absolutedouble.co.uk/trace/}, 2018.
\newblock Accessed Jan.~12, 2018.

\bibitem{acar2014web}
Gunes Acar, Christian Eubank, Steven Englehardt, Marc Juarez, Arvind Narayanan,
  and Claudia Diaz.
\newblock The web never forgets: Persistent tracking mechanisms in the wild.
\newblock In {\em Proceedings of the 2014 ACM SIGSAC Conference on Computer and
  Communications Security}, pages 674--689. ACM, 2014.

\bibitem{acar2013fpdetective}
Gunes Acar, Marc Juarez, Nick Nikiforakis, Claudia Diaz, Seda G{\"u}rses, Frank
  Piessens, and Bart Preneel.
\newblock Fpdetective: dusting the web for fingerprinters.
\newblock In {\em Proceedings of the 2013 ACM SIGSAC conference on Computer \&
  communications security}, pages 1129--1140. ACM, 2013.

\bibitem{acar18tbselenium-github}
Gunes Acar and Marc Juarez~(mjuarezm).
\newblock tor-browser-selenium: A {P}ython library to automate {T}or {B}rowser
  with {S}elenium.
\newblock The webfp/tor-browser-selenium project on GitHub:
  \url{https://github.com/webfp/tor-browser-selenium}, May 2018.

\bibitem{acar18tbselenium-pypi}
Gunes Acar~(gacar).
\newblock tbselenium: {T}or {B}rowser automation with {S}elenium.
\newblock PyPi project: \url{https://pypi.org/project/tbselenium/}, March 2018.

\bibitem{ghostwords18pull}
{Alexei (ghostwords)}.
\newblock Support {navigator.doNotTrack}.
\newblock Pull request \#1861 for the EFForg/privacybadger project on GitHub:
  \url{https://github.com/EFForg/privacybadger/pull/1861}, July 2018.

\bibitem{scriptsafe}
Andrew.
\newblock Scriptsafe: andryou.
\newblock \url{https://www.andryou.com/scriptsafe/}, 2017.
\newblock Accessed Dec.~25, 2017.

\bibitem{anonymous18torblogcomment}
Anonymous.
\newblock Comment 276687 on ``new release: {T}or {B}rowser 8.0a10''.
\newblock Tor Blog:
  \url{https://blog.torproject.org/comment/276424\#comment-276424}, August
  2018.
\newblock See responses as well.

\bibitem{canvasfingerprintblock}
appodrome.net.
\newblock {CanvasFingerprintBlock: Chrome Web Store}.
\newblock
  \url{https://chrome.google.com/webstore/detail/canvasfingerprintblock/ipmjngkmngdcdpmgmiebdmfbkcecdndc?hl=en},
  2017.
\newblock Accessed Dec.~25, 2017.

\bibitem{bravebrowser}
Brave Browser.
\newblock Fingerprint protection mode.
\newblock
  \url{https://github.com/brave/browser-laptop/wiki/Fingerprinting-Protection-Mode},
  2017.
\newblock Accessed Dec. 19, 2017.

\bibitem{cao2017cross}
Yinzhi Cao, Song Li, and Erik Wijmans.
\newblock (cross-)browser fingerprinting via os and hardware level features.
\newblock In {\em 24th Annual Network and Distributed System Security
  Symposium{NDSS}}, 2017.
\newblock
  \url{http://www.yinzhicao.org/TrackingFree/crossbrowsertracking_NDSS17.pdf}.

\bibitem{tadatitam18muon}
Amit Datta~(tadatitam).
\newblock Accept-language header has only default locale, not list of
  languages.
\newblock Issue \#429 of the Brave/Muon bug tracker on GitHub:
  \url{https://github.com/brave/muon/issues/429}, January 2018.

\bibitem{tadatitam18laptop}
Amit Datta~(tadatitam).
\newblock Fingerprinting: {B}rave's headers, plugins different from {C}hrome.
\newblock Issue \#12479 of the Brave/Browser-laptop bug tracker on GitHub:
  \url{https://github.com/brave/browser-laptop/issues/12479}, January 2018.

\bibitem{tadatitam18privacy-badger}
Amit Datta~(tadatitam).
\newblock {P}rivacy {B}adger does not set the {doNotTrack} variable in
  {JavaScript's} navigator object.
\newblock Issue \#1835 of the EFForg/PrivacyBadger bug tracker on GitHub:
  \url{https://github.com/EFForg/privacybadger/issues/1835}, January 2018.

\bibitem{chrome-plugins}
Chrome: Developer.
\newblock {NPAPI P}lugins.
\newblock \url{https://developer.chrome.com/apps/npapi}, 2018.
\newblock Accessed Jan.~12, 2018.

\bibitem{disconnect}
Disconnect.
\newblock Disconnect.
\newblock \url{https://disconnect.me}, 2017.
\newblock Accessed Jan.~12, 2017.

\bibitem{eckersley2010unique}
Peter Eckersley.
\newblock How unique is your web browser?
\newblock In {\em Privacy Enhancing Technologies}, volume 6205, pages 1--18.
  Springer, 2010.

\bibitem{panopticlick}
{Electronic Frontier Foundation}.
\newblock Panopticlick.
\newblock \url{https://panopticlick.eff.org}, 2017.
\newblock Accessed Dec 12, 2017.

\bibitem{privacybadger}
{Electronic Frontier Foundation}.
\newblock Privacy {B}adger.
\newblock \url{https://www.eff.org/privacybadger}, 2017.
\newblock Accessed Jan.~13, 2017.

\bibitem{englehardt2016online}
Steven Englehardt and Arvind Narayanan.
\newblock Online tracking: A 1-million-site measurement and analysis.
\newblock In {\em Proceedings of the 2016 ACM SIGSAC Conference on Computer and
  Communications Security}, pages 1388--1401. ACM, 2016.

\bibitem{adblockplus}
eyeo GmbH.
\newblock {Adblock {P}lus: Surf the web without annoying ads!}
\newblock \url{https://adblockplus.org}, 2017.
\newblock Accessed Dec.~27, 2017.

\bibitem{faizkhademi2015fpguard}
Amin FaizKhademi, Mohammad Zulkernine, and Komminist Weldemariam.
\newblock Fpguard: Detection and prevention of browser fingerprinting.
\newblock In {\em IFIP Annual Conference on Data and Applications Security and
  Privacy}, pages 293--308. Springer, 2015.

\bibitem{fifield2015fingerprinting}
David Fifield and Serge Egelman.
\newblock Fingerprinting web users through font metrics.
\newblock In {\em International Conference on Financial Cryptography and Data
  Security}, pages 107--124. Springer, 2015.

\bibitem{totalspoof}
fonk.
\newblock {TotalSpoof} add-on homepage.
\newblock \url{http://fonk.wz.cz/totalspoof}, 2017.
\newblock Accessed Dec.~25, 2017.

\bibitem{ghostery}
Cliqz~International GmbH.
\newblock Ghostery makes the web cleaner, faster and safer!
\newblock \url{https://www.ghostery.com}, 2017.
\newblock Accessed Dec.~27, 2017.

\bibitem{google17chorem-web-store}
Google.
\newblock Chrome web store.
\newblock \url{https://chrome.google.com/webstore/category/extensions},
  December 2017.

\bibitem{gulyas18wpes}
G\'abor~Gy\"orgy Guly\'as, Doli\`ere~Francis Som\'e, Nataliia Bielova, and
  Claude Castelluccia.
\newblock To extend or not to extend: On the uniqueness of browser extensions
  and web logins.
\newblock In {\em Proceedings of the 2018 Workshop on Privacy in the Electronic
  Society (WPES'18)}, pages 14--27, New York, NY, USA, 2018. ACM.
\newblock \url{http://doi.acm.org/10.1145/3267323.3268959}, \url
  {http://dx.doi.org/10.1145/3267323.3268959}
  {\path{doi:10.1145/3267323.3268959}}.

\bibitem{hill2015ublock}
Raymond Hill.
\newblock u{B}lock and others: Blocking ads, trackers, malwares.
\newblock
  \url{https://github.com/gorhill/uBlock/wiki/uBlock-and-others\%3A-Blocking-ads\%2C-trackers\%2C-malwares},
  May 2015.
\newblock Accessed July~5, 2017.

\bibitem{ublockorigin}
Raymond Hill.
\newblock u{B}lock {O}rigin: An efficient blocker for {C}hromium and {F}irefox.
\newblock \url{https://github.com/gorhill/uBlock}, 2017.
\newblock Accessed Dec.~27, 2017.

\bibitem{ikram2017towards}
Muhammad Ikram, Hassan~Jameel Asghar, Mohamed~Ali Kaafar, Anirban Mahanti, and
  Balachandar Krishnamurthy.
\newblock Towards seamless tracking-free web: Improved detection of trackers
  via one-class learning.
\newblock {\em Proceedings on Privacy Enhancing Technologies}, 2017(1):79--99,
  2017.

\bibitem{noscript}
InformAction.
\newblock {No{S}cript: Java{S}cript/{J}ava/{F}lash blocker for a safer
  {F}irefox experience!}
\newblock \url{https://noscript.net}, 2017.
\newblock Accessed Dec.~27, 2017.

\bibitem{canvasblocker}
kkapsner.
\newblock {CanvasBlocker}: A {F}irefox plugin to block the canvas-{API}.
\newblock \url{https://github.com/kkapsner/CanvasBlocker/}, 2017.
\newblock Accessed Dec.~25, 2017.

\bibitem{kontaxis2015tracking}
Georgios Kontaxis and Monica Chew.
\newblock Tracking protection in {F}irefox for privacy and performance.
\newblock {\em arXiv preprint arXiv:1506.04104}, 2015.

\bibitem{krishnamurthy2006generating}
Balachander Krishnamurthy and Craig~E Wills.
\newblock Generating a privacy footprint on the internet.
\newblock In {\em Proceedings of the 6th ACM SIGCOMM conference on Internet
  measurement}, pages 65--70. ACM, 2006.

\bibitem{fpcentral}
Pierre Laperdrix.
\newblock Fingerprint central.
\newblock \url{https://fpcentral.irisa.fr/}, 2017.
\newblock Accessed Oct 31, 2017.

\bibitem{laperdrix2017fprandom}
Pierre Laperdrix, Benoit Baudry, and Vikas Mishra.
\newblock Fprandom: Randomizing core browser objects to break advanced device
  fingerprinting techniques.
\newblock In {\em 9th International Symposium on Engineering Secure Software
  and Systems (ESSoS 2017)}, 2017.

\bibitem{laperdrix2015mitigating}
Pierre Laperdrix, Walter Rudametkin, and Benoit Baudry.
\newblock Mitigating browser fingerprint tracking: multi-level reconfiguration
  and diversification.
\newblock In {\em Proceedings of the 10th International Symposium on Software
  Engineering for Adaptive and Self-Managing Systems}, pages 98--108. IEEE
  Press, 2015.

\bibitem{laperdrix2016beauty}
Pierre Laperdrix, Walter Rudametkin, and Benoit Baudry.
\newblock Beauty and the beast: Diverting modern web browsers to build unique
  browser fingerprints.
\newblock In {\em Security and Privacy (SP), 2016 IEEE Symposium on}, pages
  878--894. IEEE, 2016.

\bibitem{leon2012johnny}
Pedro Leon, Blase Ur, Richard Shay, Yang Wang, Rebecca Balebako, and Lorrie
  Cranor.
\newblock Why johnny can't opt out: a usability evaluation of tools to limit
  online behavioral advertising.
\newblock In {\em Proceedings of the SIGCHI Conference on Human Factors in
  Computing Systems}, pages 589--598. ACM, 2012.

\bibitem{mayer2012third}
Jonathan~R Mayer and John~C Mitchell.
\newblock Third-party web tracking: Policy and technology.
\newblock In {\em Security and Privacy (SP), 2012 IEEE Symposium on}, pages
  413--427. IEEE, 2012.

\bibitem{blender}
meh.
\newblock Blender: Blend in the crowd by faking to be the most common {F}irefox
  browser version, operating system and other stuff.
\newblock \url{https://github.com/meh/blender}, 2017.
\newblock Accessed Dec.~25, 2017.

\bibitem{merzdovnik2017block}
Georg Merzdovnik, Markus Huber, Damjan Buhov, Nick Nikiforakis, Sebastian
  Neuner, Martin Schmiedecker, and Edgar Weippl.
\newblock Block me if you can: A large-scale study of tracker-blocking tools.
\newblock In {\em Proceedings of the 2nd IEEE European Symposium on Security
  and Privacy (IEEE EuroS\&P)}, 2017.

\bibitem{mowery2012pixel}
Keaton Mowery and Hovav Shacham.
\newblock Pixel perfect: Fingerprinting canvas in {HTML}5.
\newblock {\em Proceedings of W2SP}, pages 1--12, 2012.

\bibitem{mozilla17add-ons}
Mozilla.
\newblock Firefox add-ons.
\newblock \url{https://addons.mozilla.org/en-US/firefox/}, December 2017.

\bibitem{multiloginapp}
Multiloginapp.
\newblock How canvas fingerprint blockers make you easily trackable.
\newblock
  \url{https://multiloginapp.com/how-canvas-fingerprint-blockers-make-you-easily-trackable/},
  2017.
\newblock Accessed Dec 19, 2017.

\bibitem{glove}
Net-Comet.
\newblock {Glove: Chrome Web Store}.
\newblock
  \url{https://chrome.google.com/webstore/detail/glove/abdgoalibdacpnmknnpkgnfllphboefb?hl=en},
  2017.
\newblock Accessed Dec.~25, 2017.

\bibitem{nikiforakis2015privaricator}
Nick Nikiforakis, Wouter Joosen, and Benjamin Livshits.
\newblock Privaricator: Deceiving fingerprinters with little white lies.
\newblock In {\em Proceedings of the 24th International Conference on World
  Wide Web}, pages 820--830. International World Wide Web Conferences Steering
  Committee, 2015.

\bibitem{stopfingerprinting}
NiklasG.
\newblock Stop {F}ingerprinting: Add-ons for {F}irefox.
\newblock
  \url{https://addons.mozilla.org/en-US/firefox/addon/stop-fingerprinting/},
  2017.
\newblock Accessed Dec.~25, 2017.

\bibitem{paninski2003estimation}
Liam Paninski.
\newblock Estimation of entropy and mutual information.
\newblock {\em Neural computation}, 15(6):1191--1253, 2003.

\bibitem{torprivacy}
Mike Perry, Erinn Clark, Steven Murdoch, and Georg Koppen.
\newblock The design and implementation of the tor browser.
\newblock
  \url{https://www.torproject.org/projects/torbrowser/design/\#privacy}, 2017.
\newblock Accessed Jul 21, 2017.

\bibitem{blendin}
Re\c{s}at.
\newblock {Blend In}: Add-ons for {F}irefox.
\newblock \url{https://addons.mozilla.org/en-US/firefox/addon/blend-in/}, 2017.
\newblock Accessed Dec.~25, 2017.

\bibitem{roesner2012detecting}
Franziska Roesner, Tadayoshi Kohno, and David Wetherall.
\newblock Detecting and defending against third-party tracking on the web.
\newblock In {\em Proceedings of the 9th USENIX Conference on Networked Systems
  Design and Implementation}, NSDI'12, pages 12--12, Berkeley, CA, USA, 2012.
  USENIX Association.
\newblock \url{http://dl.acm.org/citation.cfm?id=2228298.2228315}.

\bibitem{nee}
Samy Sadi.
\newblock No {E}numerable {E}xtensions: {F}irefox addon that lets you hide
  installed extensions and avoid being fingerprinted based on them.
\newblock \url{https://github.com/samysadi/no-enumerable-extensions}, 2017.
\newblock Accessed Jan.~13, 2017.

\bibitem{salunke2014selenium}
Sagar~Shivaji Salunke.
\newblock {\em Selenium Webdriver in Python: Learn with Examples}.
\newblock CreateSpace Independent Publishing Platform, USA, 1st edition, 2014.

\bibitem{rola2017extension}
Iskander Sanchez-Rola, Igor Santos, and Davide Balzarotti.
\newblock Extension breakdown: Security analysis of browsers extension
  resources control policies.
\newblock In {\em 26th {USENIX} Security Symposium ({USENIX} Security 17)},
  pages 679--694, Vancouver, BC, 2017. {USENIX} Association.
\newblock
  \url{https://www.usenix.org/conference/usenixsecurity17/technical-sessions/presentation/sanchez-rola}.

\bibitem{privacyextension}
Martin Springwald.
\newblock {P}rivacy-{E}xtension-{C}hrome: Provides privacy for {C}hrome.
\newblock \url{https://github.com/marspr/privacy-extension-chrome}, 2017.
\newblock Accessed Dec.~25, 2017.

\bibitem{starov2017xhound}
Oleksii Starov and Nick Nikiforakis.
\newblock Xhound: Quantifying the fingerprintability of browser extensions.
\newblock In {\em Security and Privacy (SP), 2017 IEEE Symposium on}, pages
  941--956. IEEE, 2017.

\bibitem{statcounter}
StatCounter.
\newblock Stat{C}ounter global stats.
\newblock \url{http://gs.statcounter.com/}, 2018.
\newblock Accessed Feb.~12, 2018.

\bibitem{trackingprotection}
Mozilla Support.
\newblock Tracking protection.
\newblock \url{https://support.mozilla.org/en-US/kb/tracking-protection}, 2017.
\newblock Accessed Dec.~27, 2017.

\bibitem{firefox-plugins}
Mozilla Support.
\newblock Why do {J}ava, {S}ilverlight, {A}dobe {A}crobat and other plugins no
  longer work?
\newblock \url{https://support.mozilla.org/en-US/kb/npapi-plugins}, 2018.
\newblock Accessed Jan.~12, 2018.

\bibitem{tor17users}
{The Tor Project}.
\newblock Users.
\newblock Tor Metrics page:
  \url{https://metrics.torproject.org/userstats-relay-country.html}, December
  2017.

\bibitem{torres2015fp}
Christof~Ferreira Torres, Hugo Jonker, and Sjouke Mauw.
\newblock Fp-block: usable web privacy by controlling browser fingerprinting.
\newblock In {\em European Symposium on Research in Computer Security}, pages
  3--19. Springer, 2015.

\bibitem{ulmer10blog}
Hamilton Ulmer.
\newblock Browsing sessions.
\newblock Mozilla's Blog of Metrics:
  \url{https://blog.mozilla.org/metrics/2010/12/22/browsing-sessions/},
  December 2010.

\bibitem{vastel18usenix}
Antoine Vastel, Pierre Laperdrix, Walter Rudametkin, and Romain Rouvoy.
\newblock {Fp-Scanner}: The privacy implications of browser fingerprint
  inconsistencies.
\newblock In {\em 27th {USENIX} Security Symposium ({USENIX} Security 18)},
  pages 135--150, Baltimore, MD, 2018. {USENIX} Association.
\newblock
  \url{https://www.usenix.org/conference/usenixsecurity18/presentation/vastel}.

\bibitem{watson2008virtualbox}
Jon Watson.
\newblock Virtualbox: Bits and bytes masquerading as machines.
\newblock {\em Linux J.}, 2008(166), February 2008.
\newblock \url{http://dl.acm.org/citation.cfm?id=1344209.1344210}.

\bibitem{yen2012host}
Ting-Fang Yen, Yinglian Xie, Fang Yu, Roger~Peng Yu, and Martin Abadi.
\newblock Host fingerprinting and tracking on the web: Privacy and security
  implications.
\newblock In {\em NDSS}, 2012.

\bibitem{zimmeck2017privacy}
Sebastian Zimmeck, Jie~S Li, Hyungtae Kim, Steven~M Bellovin, and Tony Jebara.
\newblock A privacy analysis of cross-device tracking.
\newblock In {\em 26th USENIX Security Symposium USENIX Security 17)}, pages
  1391--1408. USENIX Association, 2017.

\end{thebibliography}

\appendix
\section{Statistical Analaysis}
\label{app:stats}

The statistical analysis used in this last step 
of the analysis engine to determine the likelihood of finding
masking depends upon the
number of different values for the attribute found across the base
browsing platforms.
As this number increases, higher-confidence (lower-valued) thresholds
$\alpha$ can be achieved.
Thus, inconclusive results can be
avoided by using a rich set of platforms.
However, this is not required for avoiding false claims
of not doing impactful masking, which is controlled by $\alpha$ alone (with
$f$ quantifying \emph{impactful}).

In more detail, we use the geometric distribution with $f$ as the
success probability.
If this probability is less than $\alpha$, then \ae{} reports
that the attribute is probably not $f$-impactfully partially standardized.
We use $0.1$ for $\alpha$ and $0.75$ for $f$, but these are
adjustable.

This approach is an estimation in two senses.
First, for attributes with a finite number of values, 
the hypergeometric distribution would give a
more accurate probability of seeing at least one
standardized value, but would require knowing the number of
possible values.
The geometric distribution underestimates this probability, making
\ae{} conservative in ruling out standardization, that is, this
estimation does not increase the rate of false claims of not doing
impactful masking.

Second, using these distributions assume that the test
attributes are drawn uniformly at random.
We instead craft them to be extreme values in hopes of triggering
standardization away from outlying values.
While this makes computing the exact probability of finding standardization
impossible, it should improve the odds of doing so except for
pathologically behaving PETs.

\end{document}